\documentclass[12pt,preprint]{aastex}

\slugcomment{submitted to {\it The Astronomical Journal 03 SEP 2014}}

\shorttitle{Variability of Red Dwarfs}
\shortauthors{Hosey et al.}

\usepackage{fixltx2e}
\usepackage{float}

\begin{document}

\title{The Solar Neighborhood XXXVI. The Long-Term Photometric
  Variability of Nearby Red Dwarfs in the $VRI$ Optical Bands}

\author{}

\author{Altonio D.~Hosey\altaffilmark{1},Todd J.~Henry\altaffilmark{1}}
\affil{RECONS Institute, Chambersburg, PA 17201}
\email{altoniohosey@gmail.com, toddhenry28@gmail.com}

\author{Wei-Chun Jao\altaffilmark{1}, Sergio B. Dieterich\altaffilmark{1}, Jennifer G.~Winters\altaffilmark{1}}
\affil{Department of Physics and Astronomy, Georgia State University, Atlanta, GA 30302-4106} 
\email{jao@astro.gsu.edu, dieterich@astro.gsu.edu, winters@astro.gsu.edu}

\author{John C.~Lurie\altaffilmark{1}}
\affil{Department of Astronomy, University of Washington, Seattle, WA 98195}
\email{lurie@uw.edu}

\author{Adric R.~Riedel\altaffilmark{1}}
\affil{Department of Astrophysics, American Museum of Natural History, New York, NY 10034}
\email{adric.riedel@gmail.com}

\and

\author{John P. Subasavage\altaffilmark{1}}
\affil{United States Naval Observatory, Flagstaff, AZ, 86001}
\email{jsubasavage@nofs.navy.mil}

\author{}
\affil{}
\email{}

\altaffiltext{1}{Visiting Astronomer, Cerro Tololo Inter-american
  Observatory.  CTIO is operated by AURA, Inc.~under contract to the
  National Science Foundation.}

\begin{abstract}

We present an analysis of long-term photometric variability for nearby
red dwarf stars at optical wavelengths.  The sample consists of 264 M
dwarfs south of Dec. = $+$30 with $V-K$ = 3.96--9.16 and $M_V$
$\approx$ 10--20, corresponding to spectral types M2V--M8V, most of
which are within 25 pc.  The stars have been observed in the $VRI$
filters for $\sim$4--14 years at the CTIO/SMARTS 0.9m telescope.  Of
the 238 red dwarfs within 25 pc, we find that only $\sim$8\% are
photometrically variable by at least 20 mmag ($\sim$2\%) in the $VRI$
bands.  Only four stars have been found to vary by more than 50 mmag,
including GJ 1207 at 8.6 pc that experienced a single extraordinary
flare, and GJ 2006A, TWA 8A, and TWA 8B, which are all young stars
beyond 25 pc linked to moving groups.  We find that high variability
at optical wavelengths over the long-term can in fact be used to
identify young stars.  Overall, however, the fluxes of most red dwarfs
at optical wavelengths are steady to a few percent over the long term.

The low overall rate of photometric variability for red dwarfs is
consistent with results found in previous work on similar stars on
shorter timescales, with the body of work indicating that most red
dwarfs are only mildly variable.  As expected, we find that the degree
of photometric variability is greater in the $V$ band than in the $R$
or $I$ bands, but we do not find any obvious trends in variability
over the long term with red dwarf luminosity or temperature.  We
highlight 17 stars that show long-term changes in brightness,
sometimes because of flaring activity or spots, and sometimes because
of stellar cycles similar to our Sun's solar cycle.  Remarkably, two
targets show brightnesses that monotonically increase (G 169-029) or
decrease (WT 460AB) by several percent over a decade.  We also provide
long-term variability measurements for seven M dwarfs within 25 pc
that host exoplanets, none of which vary by more than 20 mmag.  Both
as a population, and for the specific red dwarfs with exoplanets
observed here, photometric variability is therefore often not a
concern for planetary environments, at least at the optical
wavelengths where they emit much of their light.

\end{abstract}


\maketitle

\eject

\section{Introduction}
\label{sec:intro}

Red dwarfs comprise the majority of stars in the Galaxy, accounting
for roughly three-quarters of all stars in the solar neighborhood
\citep{2006AJ....132.2360H}.  As the lowest luminosity
hydrogen-burning stars, red dwarfs are long-lasting, stable energy
producers, making them ideal candidates for planetary environments.
One aspect of red dwarfs that may prove crucial to the potential of
life on any orbiting planets is the consistency in the flux provided
by the stars, i.e., the stellar variability.  In this paper we
investigate the long-term photometric variability at optical
wavelengths of a large sample of red dwarfs.

The RECONS\footnote{\it www.recons.org} (REsearch Consortium On Nearby
Stars) team has gathered photometry in the $VRI$ optical bands for
more than 1000 nearby red dwarfs since 1999 (Winters et al.~2011;
2014).  In this paper we focus on the long-term variability of 264 red
dwarfs (generally within 25 pc) for which we are determining
trigonometric parallaxes and searching for low mass companions as part
of our astrometry/photometry effort carried out at the CTIO/SMARTS
0.9m.  These stars typically have been observed on 10--30 nights each,
at $\sim$5 frames/night, spread over 4--14 years.  This provides a
rich dataset of more than 24,000 observations for the 264 stars, or
nearly 100 individual frames/star.  Our goals in this paper are to
determine the fraction of red dwarfs photometrically variable at
optical wavelengths over long time periods, to explore the dependence
of variability on their luminosities and temperatures, and to identify
individual stars exhibiting long-term stellar cycles.  We also briefly
address the effects of variability on the fluxes received by
exoplanets, given recent discoveries of both gas giant and
super-terrestrial exoplanets orbiting nearby red dwarfs.

Previous efforts have examined samples of M dwarfs at optical
wavelengths, often in smaller numbers or over shorter time periods.
In his classic study, \citet{1994AJ....107.1135W} reported photometric
measurements of 43 stars over an 11 year period.
\citet{2011AJ....141..117J} reported the first study of variability in
red subdwarfs, including 130 stars from the RECONS program discussed
here, observed for $\sim$3--9 years.  \citet{2010MNRAS.403.1949K}
provided photometry for over 700 stars, including many M dwarfs
observed at multiple epochs.  Transit searches have yielded rotation
periods for field K and M dwarfs via HATNet
\citep{2011AJ....141..166H} and MEarth \citep{2011ApJ...727...56I} in
datasets up to a few years in length that also provide guidance on
overall variability rates.  \citet {2012A&A...541A...9G} and
\citet{2013ApJ...764....3R} examined spectroscopic data for 27 stars
using the NaI feature and for 93 stars using the H$\alpha$ feature,
respectively, to reveal long-term changes.  Space-based efforts using
Kepler data over a few months provide results for the variability of M
dwarfs in the Kepler bandpass \citep{2011AJ....141..108C}, and
investigate their flare rates and intensities
\citep{2011AJ....141...50W, 2014ApJ...797..121H}.  These results and
ours will be discussed in more detail in $\S$8 to provide a portrait
of the dominant stellar component of our Galaxy, the red dwarfs.

\section{Sample}
\label{sec:sample}

For this study, we focus on a sample of 264 red dwarfs south of Dec. =
$+$30 observed at the CTIO/SMARTS 0.9m, some starting as long ago as
1999.  The sample is listed in Table 1, with stars falling in the
ranges of $V-K$ = 3.96--9.16 and $M_V$ $\approx$ 10--20, corresponding
to spectral types M2V--M8V.  Known cool subdwarfs have been removed,
as they have been previously discussed in detail in
\citet{2011AJ....141..117J}.  Multiple systems with separations in the
range 1--3\arcsec~have been omitted because even if two sources can be
seen in some images, accurate photometry for each source cannot always
be determined, given variable seeing.  Known multiples with
separations less than 1\arcsec~are treated as single sources and noted
with component letters in Table 1, e.g.,~AB, or ABC.  Multiples
separated by more than 3\arcsec~can be measured separately for
photometry and are given individual entries.  The names are listed in
Column 1 and the letter ``Y'' is given in Column 2 if the target is
reported to be a young star, as described in
\citet{2014AJ....147...85R}.  After the coordinates in Columns 3 and
4, we provide the $VRI$ magnitudes and references from our program in
Columns 5--8, followed by the $K_s$ magnitude from 2MASS and the
$V-K_s$ color in Columns 9--10.  In Columns 11--16, we list the filter
used for the long-term astrometric observations, the photometric
variability in that filter calculated using methods described in
$\S4$, and various metrics of the duration of the observations,
including the first and last epochs in each data series, the time
coverage, the number of nights on which observations were taken, and
the total number of frames (typically $\sim$5 frames per night).  The
final columns list the parallax values and errors for stars in our
sample for which we have published parallaxes to date, and the
references.  To determine $M_V$ values used to analyze the sample, the
published parallaxes are used, as well as unpublished values also from
the RECONS program for which final values will be presented in future
papers in this series.

As shown in Figure 1, the median time coverage for stars in the sample
is 7.9 years, with a broad distribution that stretches from 3.8 years
to 14.0 years of observations for stars that were targeted at the
beginning of the program.  The primary focus of the program at the
0.9m involves the discovery and characterization of nearby stars, with
most stars within 25 pc, but the set of stars presented here is by no
means a complete, volume-limited sample.  Target stars are usually
selected based upon their high proper motions
\citep{2004AJ....128..437H, 2005AJ....129..413S, 2005AJ....130.1658S,
2007AJ....133.2898F, 2011AJ....142...10B, 2011AJ....142...92B,
2011AJ....142..104R} or photometric attributes
\citep{2004AJ....128.2460H, 2011AJ....141...21W} that indicate they
are likely to be closer than 25 pc.  However, there are two groups of
red dwarfs on the program that were eventually found to be at
significantly different distances than anticipated.  Cool subdwarfs
\citep{2005AJ....129.1954J, 2011AJ....141..117J} fall below the main
sequence in the H-R diagram, and are closer than their photometric
distance estimates imply.  Young red dwarfs
\citep{2011AJ....142..104R, 2014AJ....147...85R} lie above the main
sequence and are consequently further than their photometric distance
estimates.  Thus, the best assessment of variability rates in a census
of stars includes a detailed analysis of their distances, to prevent
over or underrepresentation of subsamples of stars that may bias the
statistics.  Of the 264 stars in the sample discussed here, 238 are
known to be within 25 pc based on our trigonometric parallaxes.
Below, we analyze both the entire list of stars observed and the
restricted sample of only those within 25 pc.


\section{Observations and Data Reduction}
\label{sec:observations}

RECONS has been using the CTIO/SMARTS 0.9m telescope for astrometric
and photometric observations since 1999, first as an NOAO Surveys
Program, and since 2003 under the auspices of the SMARTS Consortium.
The telescope is equipped with a 2048 x 2048 Tektronix CCD camera that
is rarely moved, which provides the long-term astrometric stability
needed for parallax work \citep{2014AJ....147...94D,
  2006AJ....132.2360H, 2005AJ....129.1954J, 2011AJ....141..117J,
  2014AJ....147..21J, 2010AJ....140..897R, 2011AJ....142..104R,
  2014AJ....147...85R, 2009AJ....137.4547S} as well as a stable
photometric platform.  Images taken during the program are used here
to investigate the photometric variability of the nearby M dwarfs that
have been targeted for parallax and proper motion measurements.
Observations are made using the central quarter of the chip, which
provides a 6\farcm8 square field of view and pixels 401
milliarcseconds in size.  Parallax frames are taken in the $V$, $R$,
and $I$ filters\footnote{The central wavelengths for the $V_J$,
  $R_{KC}$, and $I_{KC}$ filters used in this study are 5438/5475,
  6425, and 8075 \AA, respectively.  The two $V$ filters used during
  this study have been found to be photometrically identical at a
  level better than the 7 mmag minimum level of our variability
  sensitivity, as there are no offsets observed in the more than 100
  stars observed in the $V$ band \citep{2011AJ....141..117J}.  The
  subscript ``J'' indicates Johnson, ``KC'' indicates Kron-Cousins
  (usually known as Cousins), and are hereafter omitted.}  with
magnitudes ranging from 9 to 20.

For astrometry, five images of each star are typically taken per
night, usually within 30 minutes of transit.  The target star is
positioned in the field so that 5--10 reference stars, normally
fainter by 1--4 magnitudes, surround the target.  These stars
constitute a reference grid for the astrometric reductions, and are
also used for the photometric variability study described here.
Additional details of the observations can be found in
\citet{2005AJ....129.1954J}.  Exposure times usually range from
10--300 sec, with a few of the faintest stars requiring 600-900 sec
integrations.

Data reduction includes calibration using flatfield and bias frames
that are taken nightly.  Each science frame is then manually checked
for saturation of the target star or any reference stars.  A frame
with a saturated target star is discarded; individual reference stars
are discarded if saturated, but the frame is used if sufficient
reference stars remain available for reliable relative photometry.

$VRI$ photometry from our program is given for the sample stars in
Table 1.  Details of the photometry observations and reductions can be
found in \citet{2005AJ....129.1954J} and \citet{2011AJ....141...21W}.
Briefly, calibration frames are taken nightly, and standard stars
selected from \citet{1990A&AS...83..357B},
\citet{1982PASP...94..244G}, and \citet{1992AJ....104..372L,
  2007AJ....133.2502L} are observed multiple times each night in order
to derive transformation equations and extinction curves.  Apertures
14$\arcsec$~in diameter were used to determine the stellar fluxes,
except in cases where close contaminating sources needed to be
deblended, in which case smaller apertures were used and aperture
corrections were applied.  Photometric errors for the $VRI$ magnitudes
are typically 0.03 mag, measured using (usually) multiple nights of
data and taking into account extinction corrections for each night via
standard star observations.  These errors are much larger than the
relative photometry for the variability measurements discussed below,
which instead use stars in the target fields at virtually identical
airmasses.

\section{Variability Measurements}
\label{sec:variability}

Here we define photometric variability to be the standard deviation of
a star's flux, measured in milli-magnitudes (mmag), when compared to a
set of reference stars.  Once a setup frame that positions a target
star within an ensemble of reference stars has been established, the
target star's magnitude in each frame is compared to the reference
stars using the methodology outlined in \citet{2011AJ....141..117J}.
As an amendment to that methodology, we incorrectly stated in that
paper that our instrumental magnitudes were based on counts within a
defined aperture, whereas in fact they are based on Gaussian fits to
the light distribution of each source.  Briefly, we calculate
instrumental magnitudes via SExtractor by integrating an object's
pixel values within a circular Gaussian profile that is scaled to an
object's image size and shape in a frame.  The FWHM of this circle
will be the radius of the disk that contains half of the object's
flux.  We control for changes in seeing, airmass, and atmospheric
transparency in a series of exposures for an object by utilizing the
prescription discussed in \citet{1992PASP..104..435H}.


True stellar variability may be due to short-term flaring activity,
mid-term rotation with spots, or long-term changes in spot numbers,
i.e., a stellar cycle.  Variability in frames due to other causes is
identified and removed from the dataset.  Such false variability may
be caused by frames compromised by high background because they were
taken in twilight or moonlight, or by contamination by a nearby source
as the proper motion of the target star causes its position to slide
across the field.  One of the advantages to this long-term program and
its fairly high resolution images (401 milliarcsecond pixels) is that
the positions of background sources can be identified and monitored
over time relative to target stars, thereby eliminating contaminating
sources.  Finally, reference stars that do not fit the trend of
instrumental magnitude standard deviation with brightness
\citep[see][Figure 3]{2011AJ....141..117J} are removed from the
analysis, and the target star's brightness is compared to the
remaining ``quiet'' reference stars.

For the stars discussed here, we find standard deviations in the
photometric series as low as 6.4 mmag, although only a half dozen
stars have standard deviations below 7 mmag, which we adopt as the
one-sigma minimum deviation threshold.  This matches the 7--8 mmag
variability measured for three stars in \citet{2011AJ....141..117J}
with long time-series of images using two different $V$ filters.
Thus, mixing the two $V$ filters is not a serious concern because the
filters appear to be so similar that any offsets are much smaller than
the 7 mmag minimum variability threshold we can measure.  This level,
which we refer to henceforth as the detectable ``variability floor'',
is represented with horizontal solid lines at 7 mmag in Figures 2--4.
As a conservative measure of variability, we define significantly
variable stars as those with standard deviations of 20 mmag or more
relative to the chosen reference star set, but note that some stars in
the 15--20 mmag range are variable at a lower level.  This value is
measured on an absolute scale, so that $-$20 mmag and $+$20 mmag
differences would both be regarded as variability by 20 mmag.

\section{Results}
\label{sec:results}

In Figures 2--4 we show photometric variability as functions of
apparent $VRI$ magnitude, $V-K_s$ color, and $M_V$.  There are 114
stars observed in the $V$ filter, 81 in the $R$ filter, and 69 in the
$I$ filter.

In the three panels of Figure 2, there are no obvious trends in
variability with apparent magnitude, but there are more stars variable
by at least 20 mmag in $V$ than in $R$ or $I$.  The intrinsically
faintest, reddest, stars are typically observed in the $I$ filter to
boost S/N, and as discussed in $\S$6, among the three filters this is
where we find the least variability, as expected.  Our variability
floor of 7 mmag is the same in all three filters and at all
brightnesses because we integrate for longer times on fainter targets
to boost S/N for the astrometry studies, with the consequence that our
photometric errors are not a strong function of brightness
\citep{2011AJ....141...21W}.  We therefore conclude that systematic
errors concerning our lower limit of variability detection have been
ameliorated, and that the overall observational thresholds for
variability detection are consistent among the three filters and at
various target brightnesses.

The three panels of Figure 3 show perhaps a subtle trend of long-term
variability with $V-K$ color, which corresponds to temperature, with
redder M dwarfs perhaps a bit more variable in $V$ and $R$, but not in
$I$, than early-type M dwarfs.  However, this trend is weak, if
present at all, and only with a much larger sample might the trend be
confirmed.  The three panels of Figure 4 are similar to those in
Figure 3, now using absolute magnitudes to explore trends in
luminosity rather than temperature.  The results are predictably
similar to Figure 3 because luminosity is linked to temperature for
main sequence stars.


\section{Discussion}
\label{sec:discussion}

Inspection of Figures 2--4 reveals a distribution of long-term
variability in red dwarfs that we divide into two populations: (1)
clearly variable stars that change brightness by more than 20 mmag,
with a few stars highly variable at more than 50 mmag, and (2) those
that are relatively quiescent with variability less than 20 mmag.  The
division at 20 mmag is used to compute the fractions of variable
stars, illustrated using a restricted sample of stars in the histogram
of Figure 5 (discussed below).  As illustrated in Figures 6--8, there
are some likely variable stars in the relatively quiescent group with
variability measurements of 15--20 mmag, but our goal here is to
understand what fraction of red dwarfs vary by a significant amount,
here chosen to be a threshold of 20 mmag, or $\sim$2\% in flux in the
observed bands.  In the full sample, only 24 (9\%) of the 264 stars
vary photometrically by at least 20 mmag over long timescales; thus,
$\sim$90\% of red dwarfs change in optical brightness by less than
2\%.  For the full sample, the fractions of stars that fall into the
variable/quiescent groups are 18\%/82\% in the $V$ band (114 stars),
4\%/96\% in the $R$ band (81 stars), and 3\%/97\% in the $I$ band (69
stars).

These fractions are subject to the biases of the observational
program, which targets red dwarfs possibly within 25 pc.  The primary
bias of concern is the sample of young stars that are larger and
intrinsically brighter than main sequence dwarfs, resulting in
distance estimates placing them much closer than their true distances.
There are 15 young stars, as determined by
\citet{2014AJ....147...85R}, in our sample, denoted with ``Y'' in
Column (2) of Table 1, and with encircled points in Figures 2--4.
Seven of these stars were observed in the $V$ filter and eight in the
$R$ filter.  Seven of the stars vary by more than 20 mmag (six in $V$
and one in $R$), implying that 47\% are variable stars by our
criterion, a much higher fraction than in the sample as a whole.

Ideally, we would use a complete, volume-limited, sample to measure
the fractions of red dwarfs photometrically variable over long
timescales.  Our best possible assessment with the current dataset is
to restrict the sample to the 238 stars known to be within 25 pc via
our trigonometric parallaxes, which is at least more representative of
the red dwarf population than the full sample.  Among these, the
fractions of variable/quiescent red dwarfs are 13\%/87\% in the $V$
band (106 stars), 4\%/96\% in the $R$ band (70 stars), and 3\%/97\% in
the $I$ band (62 stars).  Overall, only 19 of the 238 stars (8\%) are
variable by 20 mmag or more.  We retain six young stars in the 25 pc
sample (two observed at $V$ and four at $R$) that vary by 17--25 mmag.
All six stars vary by more than the median variability values in the
$V$ (14 mmag), $R$ (13 mmag), and $I$ filters (11 mmag).

Somewhat to our surprise, we do not see any strong trends in long-term
variability with color (Figure 3) or luminosity (Figure 4) through the
full range of red dwarfs in our sample.  From the earliest types that
have $V-K$ = 4.0, $M_V$ $\sim$ 10, spectral type M2V, and $T$ = 3600K,
to those with $V-K$ = 9.2, $M_V$ $\sim$ 20, spectral type M8V, and $T$
= 2200K, (for empirical definitions of M dwarfs, see
\citet{1994AJ....108.1437H}, \citet{1996ApJ...461L..51B},
\citet{2006AJ....132.2360H}, and \citet{2014AJ....147...94D}), there
are no obvious trends in activity with stellar type seen in any of the
$VRI$ bands.  This is a particularly revealing result --- stars that
differ by a factor of $\sim$10$^4$ in visual flux (using M$_V$) and
$\sim$160 in total luminosity do not exhibit any trends in their
photometric variability over long time periods in the bands where they
exhibit a sizeable amount of flux.  For example, M0V stars emit 46\%
of their flux in the $VRI$ bands, in contrast to M5V stars that emit
17\% in the same bands.  Furthermore, because red dwarfs lose their
radiative zones and become fully convective around $V-K$ $\approx$
4.5, $M_V$ $\approx$ 11.0 \citep{1997A&A...327.1039C}, we might also
expect to see an abrupt change in long-term photometric variability
character at that color and absolute magnitude.  We do not see such a
change in the current sample, but we have not observed many stars
bluer than $V-K$ = 4.5.

To evaluate the relative variability in the three bands, we restrict
the 25 pc sample further to include only stars with $V-K$ = 4.5--6.5,
in order to limit the effects of fainter, redder, stars in the sample
that are typically observed in the $I$ filter.  This subsample
contains a total of 180 stars, with median variability values of 14
mmag at $V$ (94 stars), 13 mmag at $R$ (66 stars), and 11 mmag at $I$
(20 stars).  These values are unchanged from the analysis including
stars of all colors within 25 pc, and we conclude that red dwarfs vary
slightly more in the $V$ and $R$ bands than in the $I$ band over
multi-year timescales, confirming what we reported in
\cite{2011AJ....141..117J}.


In Figure 5 we show the fractions of stars within 25 pc observed in
$V$ that are variable by more than 20 mmag, as a function of $M_V$.
The range is restricted to only those bins that have at least 10
stars, thereby including 97 of the 106 stars meeting the selection
criteria.  The error bars represent counting statistics and illustrate
that more stars are needed to discern if there are any trends with
luminosity --- the total of 13 stars meeting our variability threshold
is too small a sample to identify any trend.

\section{Systems Worthy of Note} 
\label{sec:systemnotes}

Here we provide details on several stars observed during the survey,
grouped by type of variability and listed alphabetically within each
group.  Example plots of variability in the various categories are
presented in Figures 6--8.  We break the stars into those showing
stellar cycles, trends, and erratic brightness changes likely due to
spots and/or flares.  We note that some of these stars do not formally
exceed our conservative 20 mmag threshold for obvious variability, but
they are highlighted here primarily because we see compelling trends
in their data sets.

In the upper left panel of Figure 6, we first show a baseline
non-variable target, {\bf SCR 1845-6357AB}, which is composed of an
M8.5V dwarf and a brown dwarf companion separated by
$\sim$1\arcsec~that does not affect the photometry.  In fact, this
target has such a low level of variability that we have adopted it as
a red photometric standard.

{\bf Stellar Cycles in GJ 831AB, GJ 1061, LHS 2397aAB, LP 467-016AB,
  and SCR 0613-2742AB:} These five targets exhibit cyclic variations
in their photometry, illustrated in Figure 6.  GJ 831AB is a
fast-orbiting binary with an orbital period of 1.93 years and
$\Delta$$V$=2.1 between the two components \citep{1997AAS...191.9302F,
  1999ApJ...512..864H}.\footnote{A possible third component mentioned
  in those papers has not been confirmed through continued work on the
  dataset, so the system is a binary, not a triple.}  We see a pattern
in our current dataset with a period of $\sim$8 yr, although there are
deviations from a clear sinusoidal cycle in 2005.  Given
$\Delta$$V$=2.1, presumably the observed variability is in the primary
component.  The variation in GJ 1061 is muted, with a tentative cycle
lasting about 8 years.  The variations in LHS 2397aAB and LP 467-016AB
appear to be robust, with periods of $\sim$4 years and $\sim$5 years,
respectively.  The most obvious cycle is that of SCR 0613-2742AB, for
which we derive a period of $\sim$0.4 years.  Because of the short
period for the cyclic behavior noted for SCR 0613-2742AB, we are
likely seeing evidence of the rotation period, while the longer term
upward trend may reflect a portion of a long-term cycle.

{\bf Photometric Trends in G 169-029, GJ 876, L 449-001AB, LHS 1610AB,
  LHS 2021, and WT 460AB:} These six targets show long-term trends in
their photometry, illustrated in Figure 7, but have not yet clearly
completed a cycle in our coverage.  Remarkably, G 169-029 (10 years)
and WT 460AB (13 years), show extraordinarily long-term trends that
have yet to invert, with overall flux changes of $\sim$5\% in $R$ and
$I$, respectively.  To our knowledge, these are the first discoveries
of changes in red dwarf fluxes lasting a decade or more.  WT 460AB is
a binary composed of two red dwarfs with spectral types M5.5V and
$\sim$L1V separated by 0\farcs5 \citep{2006A&A...460L..19M}.  The
large $\Delta$$H$=2.5 mag difference implies an even larger
$\Delta$$I$, so presumably the change in brightness is due to the M
dwarf.  The data series on L 449-001AB similarly shows a long-term
trend throughout the 5-year time series in-hand.  Additional trends
are seen in the data for G 161-071, LHS 547, and SCR 2036-3607 (not
shown here).

{\bf Spots in GJ 2006A, LP 932-083, and TWA 8A:} These targets show
distinctive variability illustrated in the left three panels of Figure
8, at levels of 77 mmag, 46 mmag, and 79 mmag, respectively.  The data
indicate that these stars change their broadband fluxes frequently,
likely because much of the stellar surface is active with flares
and/or spots.  Flares and spots are both modulated by stellar
rotation, as the active regions rotate in/out of our view, and as the
spots change over long time periods.  We do not see obvious cyclic
patterns in the brightnesses of these three stars.  The case of TWA
8AB is particularly intriguing.  This is a binary member of the TW
Hydra Association separated by 13 arcseconds, in which the A component
shows clear spot activity and for which we have observed a flare in
the B component (see below).  Additional evidence for spots is seen in
the data for G 007-034, G 131-026, GJ 1123, GJ 1284AB, GJ 2006B, LHS
2206, LHS 5094, LP 834-032, and Proxima Centauri (not shown here).

{\bf Flares in GJ 1207, LHS 6167AB, and TWA 8B:} These three stars
show clear flare events, illustrated in the right panels of Figure 8.
GJ 1207 has the largest variability measurement, 196 mmag, in the
sample, caused entirely by our observation of a flare that brightened
the star by 1.7 mag in $V$ on UT 2002 June 17
\citep{2006AJ....132.2360H}.  Omitting the flare event reduces its
variability to 18 mmag.  LHS 6167AB shows a flare with amplitude 0.2
mag on UT 2013 April 2.  TWA 8B is the second most variable star in
our sample, at 124 mmag, again caused by single 0.7 mag flare in $V$
observed on UT 2000 March 27.

Overall, there are only four red dwarfs in the sample that vary by
more than 50 mmag: GJ 1207, GJ 2006A, TWA 8A, and TWA 8B, all observed
in the $V$ filter.  The GJ 2006 (the B component varies by 36 mmag in
$V$) and TWA 8 systems are young stars, as discussed in
\citet{2014AJ....147...85R}.  GJ 1207 is not known to be a young
system, and without the single extreme flare event mentioned above, it
would fall in our quiescent group.

\section{Comparison to Previous Studies}
\label{sec:previous}

\subsection{Long-Term Results from Ground-Based Studies}

In his classic study, \citet{1994AJ....107.1135W} used a GaAs
photomultiplier tube and a similar $VRI$ filter set to that used in
our study.  Although Weis was not able to guarantee the same tube and
filter set for every run (typically seven to eight nights, which he
referred to as a season of observation), he provided valuable results
on the variability of nearby M dwarfs.  He observed 43 stars for up to
11 years in all three of the $VRI$ filters, and determined a
variability detection limit of 5 mmag for his observational technique.
Weis used $\sigma_{a}$ to denote the average mean error of a seasonal
mean magnitude and $\sigma_{b}$ to denote the dispersion of the
seasonal mean magnitudes about the overall mean.  These two quantities
correspond to short-term (over a $\sim$week) and long-term (over
years) variability measurements.  He concluded that 21 of the stars
exhibited long-term variability at the 95\% confidence level.
However, most of the variability was in the range 10-20 mmag; only 8
(19\%), 4 (9\%), and 0 (0\%) stars were variable at the 20 mmag level
over long time periods in the $V$, $R$, and $I$ bands, respectively.
Our variability measurements are similar to Weis' $\sigma_{b}$
measurement, and we find similar fractions of variability, albeit for
$\sim$six times the number of stars: 18\%, 4\%, and 3\% of our stars
vary by more 20 mmag or more at $V$, $R$, and $I$, respectively.  To
our knowledge, Weis' work is the first significant study of the
long-term variability of red dwarfs.

\citet{2011AJ....141..117J} studied a set of 22 cool subdwarfs, and
compared them to 108 main sequence red dwarfs.  Those results are
based on the same long-term astrometry/photometry taken during the
RECONS program at the 0.9m outlined above, but over a somewhat shorter
time period.  For the first time, that study revealed that red
subdwarfs are less photometrically variable in the $VRI$ bands than
their main sequence counterparts, with average variabilities for the
22 cool subdwarfs at levels of only 7 mmag at $V$ (3 stars), 8 mmag at
$R$ (13 stars), and 7 mmag at $I$ (6 stars).  This is effectively the
variability floor level of our observations, and in fact, cool
subdwarfs are likely even less variable than indicated by those
values.

Spectroscopic variability studies of M dwarfs spanning years have been
carried out by \citet{2012A&A...541A...9G} and
\citet{2013ApJ...764....3R}.  Both studies focused primarily on stars
with spectral types M0V--M4V, a bluer set of stars than our M2V--M8V
sample.  \citet{2012A&A...541A...9G} investigated 27 M dwarfs in the
HARPS radial velocity program with a median observational timespan of
5.9 years to detect correlations between long-term activity variations
and the measured radial velocities.  By using the NaI D doublet as an
activity index, they found that 14 of the stars had activity
indicative of long-term variations, although in truth, only a few of
the NaI vs.~time plots show convincing cycles.  Assuming their
selection of the five stars deemed to have reliable sinusoidal signals
in their NaI indices, their variable activity rate is 19\%.


In a comparable spectroscopic study, \citet{2013ApJ...764....3R} used
$\sim$11 years of spectra from the McDonald Observatory M Dwarf Planet
Search to reveal cyclic variations in M dwarfs that could be measured
using the H$\alpha$ line.  Among their 93 stars they identified five
exhibiting periodic signals ranging from 0.8--7.4 years, and found
eight additional stars showing long-term trends, or at least offsets
in H$\alpha$ index over time.  Their fraction of stars with detected
changes in H$\alpha$, 14\%, is similar to our fraction of stars
varying by more than 20 mmag in the $V$ band, 18\%, implying that
changes in H$\alpha$ may correlate quite well to changes in $V$ band
flux.

The single star included in \citet{2012A&A...541A...9G},
\citet{2013ApJ...764....3R}, and our study is the multi-planet system
GJ 581.  The spectroscopic studies report periodic signals in NaI
lasting 3.9 years in a 6-year dataset and in H$\alpha$ lasting 4.5
years in a 9-year dataset, respectively.  GJ 581 falls in the
quiescent set of our stars, varying by only 13 mmag in $V$ over 12.8
years.  We note that GJ 581 has the largest false alarm probability
(0.12) of the five stars with periodic signals in
\citet{2013ApJ...764....3R}, and although \citet{2012A&A...541A...9G}
state that this is the star with the highest probability that its
activity data can be fitted with a sinusoidal signal, the amplitude is
only twice the average error.  We conclude that this particular cycle
may not be real.  Nonetheless, long-term photometric monitoring of GJ
581 and the other NaI and H$\alpha$ variable stars would certainly be
a worthy project, particularly at the $\sim$1 mmag level possible with
ground-based telescopes.


\subsection{Shorter-Term Results from Ground-Based Studies}

\citet{2011AJ....141..166H} explored the optical variability of a
sample of 27560 field K and M dwarfs using data from the HATNet
transit-search survey.  Their observational series range from 45 days
to 2.5 years, with a median time span of half a year.  Thus, they are
sampling time periods suited for exploring rotation periods, rather
than the stellar cycles we are investigating.  After carefully
deleting blended sources, a concern because of HATNet's
9--14\arcsec/pixel scales, they find 1490 stars (excluding eclipsing
binaries) that vary at the 10 mmag level in the $RI$ bands, indicating
that at least $\sim$5\% vary at this level.  In their Figure 14, they
show an increasing fraction of variable stars with $V-K$ color.  For
stars having $V-K$ = 4.5--6.5, like most of those in our study, they
find 10 mmag variability rates rising from 7\% to more than 40\%,
although the error bars are significant, particularly beyond $V-K$ =
5.25 because there are 20 or fewer stars in each of those three bins.
Regardless, this large survey indicates that most M dwarfs {\it do
  not} vary by more than 1\% over many months, similar to our
conclusion for similar stars over many years.

\citet{2011ApJ...727...56I} reported rotation periods, based upon
photometric variability, in the range of a few hours up to about 5
months for 41 red dwarfs from the MEarth project.  Their study differs
from that of \citet{2011AJ....141..166H} in that they were
specifically targeting nearby stars, and using much higher resolution
images, with pixels 0.76\arcsec~in size so blending is rarely an
issue.  As with the HATnet survey, their observations are higher
cadence than ours, with typically several hundred to several thousand
observations per star over time periods of 3 months to 3 years, with
an upper sensitivity limit to periodicities at 5 months.  The
observations were made through a long pass filter at 715 nm, and
periodic variations were reported with amplitudes of 2.7 to 23.9 mmag
in their Table 1.  Their study does not examine the long-term
variability discussed here, but as with \citet{2011AJ....141..166H},
provides important complementary information.  The 41 stars with
rotation periods were extracted from an overall sample of 273 stars,
of which $\sim$80\% had at least 100 observations on 10 or more
nights, indicating that $\sim$20\% of the well-sampled stars had
determinable rotation periods.  Among these, 10 have semi-amplitudes
of at least 10 mmag, or roughly 5\%, similar to that found by
\citet{2011AJ....141..166H}, although no information was provided
about the variability rate for stars not found to have rotation
periods.

Finally, \citet{2010MNRAS.403.1949K} reported $UBVRI$ photometry for
over 700 nearby stars, primarily of spectral types K and M.  Their
study is rather different than ours, as it was an effort to provide
photometry, with typically a handful of observations per star rather
than long time series over many years.  Nonetheless, as outlined in
their Figure 3 (based on apparent $V$ magnitude), they find that most
of their sample stars vary by less than 20 mmag, similar to our
result.


\subsection{Results from the Kepler Mission}

To date, the Kepler mission has provided an opportunity to study a few
thousand M dwarfs in the northern hemisphere photometrically, although
most are not in the immediate solar neighborhood, and none are in
common with the stars studied here.  Kepler samples the light curves
of stars in a bandpass that spans 400-850 nm, every 30 minutes for the
datasets discussed here.  This single bandpass roughly corresponds to
the combined light of the $VRI$ bands we have used.

In a study of the first month of Q1 Kepler science data,
\citet{2011AJ....141..108C} extracted variability rates for periods up
to 33 days for 129,000 dwarfs, including more than 2000 identified as
M dwarfs.  There are two apparent magnitude samples examined that
provide useful statistics, but there are several {\it caveats}: the
stars are at various distances, sources are defocused so may be
unresolved multiple systems or blends with background stars, and while
significant efforts were made to separate the dwarfs and giants, no
color information for individual stars is given, so various types of M
dwarfs (and possibly other red objects) were mixed in the samples.
Using data in their Table 2, for M dwarfs having $M_{Kepler}$ =
12--14, 14 of 154 stars (9\%) had magnitude dispersions greater than
10 mmag over the 33 day period.  For $M_{Kepler}$ = 14--16, 150 of
2182 stars (7\%) varied by 10 mmag.  In addition, they find that the
overall variability fraction at the 30-minute sampling rate increases
for the M dwarfs as baselines increase from 1 to 33 days, and
therefore conclude that M dwarfs vary primarily on timescales of weeks
or longer.  Because our variability floor is 7 mmag, a direct
comparison to the results from Kepler, which measures photometry to
tens of parts per million, is problematic --- we find that 193 (73\%)
of the 264 stars in our sample vary by at least 10 mmag in one of the
three $VRI$ filters, but this large fraction is simply because 10 mmag
is very close to our variability floor.



In a second Kepler study, \citet{2011AJ....141...50W} examined $\sim$
23,000 K and M dwarfs in the same Q1 dataset as
\citet{2011AJ....141..108C}, again using the 30-minute cadence results
over a 33 day time period, with the goal of identifying flare stars.
They found 373 stars that exhibited flare activity via their
$EW_{phot}$ parameter.  They derive three characteristics of the stars
in their flare sample: (1) M dwarfs flare more frequently than K
dwarfs, but for shorter durations, (2) there is no dependence on the
flare peaks with stellar effective temperature, and (3) stars that
have larger quiescent variability have intrinsically larger flares.
Overall, the fraction of stars exhibiting flares in the Kepler dataset
is very low, less than 2\%.  This is consistent with our data, given
that we have identified only three stars that clearly flare among 264.

More recently, \citet{2014ApJ...797..121H} report flare occurrence
rates for five M dwarf systems (four singles and one close double)
monitored by Kepler in short cadence mode.  Three of the M dwarfs had
no H$\alpha$ emission and were labeled ``inactive'', yet still
produced a few energetic flares, but at much lower rates than their
active counterparts.  These three inactive M dwarfs were in flaring
states for 0.1\%, 0.6\%, and 1.5\% of the time observed, while the two
active M dwarf systems spent 27\% and 36\% of the time in a flaring
state.  Thus, chromospherically inactive M dwarfs do exhibit energetic
flares, but only rarely, which is consistent with the very few flares
we have seen in more than a decade of monitoring nearby red dwarfs.



\subsection{Summary of Results}
\label{sec:summary}

The long term efforts of \citet{1994AJ....107.1135W},
\citet{2011AJ....141..117J}, \citet{2012A&A...541A...9G}, and
\citet{2013ApJ...764....3R}, as well as our study presented here, all
indicate that only $\sim$10--20\% of red dwarfs vary by at least 2\%
in the $V$ band, or when using proxy spectral features.  At $R$ and
$I$, the variability rate is much lower.  In sum, these studies
clearly indicate that M dwarfs do exhibit long-term cycles like our
Sun, but that most M dwarfs do not vary by more than a few percent
over long timespans.

The remaining optical wavelength studies discussed here focus on
timespans suited to determining rotation periods, rather than stellar
cycles.  Those results explore timescales up to a few months in
duration, whereas our measurements are over years, yet in sum also
indicate that only a small fraction of M dwarfs exhibit flux
variability exceeding a few percent.  Although the $VRI$ bands in
which we have observed the stars in this survey may not be those in
which variability is greatest, such as at x-ray wavelengths, red
dwarfs emit much of their flux in these bands, in particular for the
earlier types.  Most of the rest of red dwarfs' flux is emitted in the
near-infrared, and the study by \citet{2012MNRAS.427.3358G} of
$\sim$9600 M dwarfs yielded only 68 stars (less than 1\%) that were
periodically variable by more than 10 mmag (their Table A1) in the $J$
band.  This is a much smaller fraction than we found at $VRI$, and
confirms the trend of less variability with increasing wavelength.

%

\section{M Dwarf Variability and Exoplanets}
\label{sec:exoplanets}

One of the motivations for this study was to evaluate the variability
of the ubiquitous red dwarfs because such variations will affect the
environments, and perhaps the habitability, of any orbiting planets.
In particular, long-term variations in the flux emitted could affect
planetary atmospheres in ways similar to how our Sun affects the
Earth's atmosphere, e.g., the cyclic influx of charged particles that
cause aurorae and expansion/contraction of the atmosphere.  Here we
discuss some of the nearest red dwarfs with detected exoplanets to
provide context for effects on planetary environments that stretch
over years.

In Table 2, we list M dwarfs within 25 pc currently reported to have
orbiting exoplanets.  This sample has been created using the
intersection of the exoplanets.org and exoplanet.eu websites as of 01
July 2014 --- stars are only included if the detected exoplanets have
been vetted and listed by both groups.  The sample is heavily skewed
to the closest M dwarfs; presumably many more planets await discovery
within 25 pc.  We list the number of reported planets and photometry
for the 17 stars, as well as our own monitoring results for seven
stars, including $VRI$ photometry, filters used for the long-term
observations, the variability levels, and the duration and number of
observations (nights, frames).  The unobserved stars in Table 2 are
not being followed because they are either too bright for our
astrometric program, requiring short exposures and consequently poor
centroids for astrometry, or are in the northern sky.

None of the seven exoplanet host stars we have observed has been found
to be variable by more than 20 mmag in the available datasets.  The
only star of note among the seven is GJ 876, which displays a possible
stellar cycle (Figure 7), growing brighter from 2004--2006, and rather
fainter in the most recent epoch available in 2013.  We conclude that
photometric variability, at least at optical wavelengths, is not a
concern for the environments of exoplanets orbiting most red dwarfs.

\section{Conclusions}
\label{sec:conclusions}

We have collected photometric observations for 264 red dwarfs over the
past 14 years, with a median duration in the coverage of 7.9 years.
We have used these images to determine long-term photometric
variability in the $VRI$ bands, and reach the following conclusions:

$\bullet$ Only $\sim$10\% of red dwarfs are variable at the 20 mmag
level over multi-year timescales in the $VRI$ photometric bands.
Thus, the impression that red dwarfs are highly active with large
changes in brightness is likely due to the notoriety of a small number
of highly variable stars, e.g. EV Lac.  Instead, most red dwarfs are
relatively quiet, at least at optical wavelengths.

$\bullet$ As expected, we find that the measured variability is lower
in the $I$ band than in the $V$ or $R$ bands.  Of the 25 stars varying
by more than 20 mmag, 20 were observed in the $V$ band.  Given that
most red dwarfs emit the majority of their flux in the $I$ band, or at
even longer wavelengths in the near-infrared where the variability is
even lower, the overall fluxes of most red dwarfs change very little.

$\bullet$ To our surprise, we find no clear trends in variability at
the 20 mmag level for M dwarfs as functions of luminosity or
temperature, at least in the current sample.  However, identifying
relatively young stars can be accomplished via long-term photometric
monitoring.

$\bullet$ We have identified several red dwarfs exhibiting long-term
photometric changes, including five stars with cyclic periods lasting
up to 8 years and nine stars with long-term trends, two of which have
been continuing for a decade.  At least 12 additional stars show
evidence of spots.

These results have a particular application to the study of habitable
environments around M dwarfs.  To first-order, planetary surface
temperatures are estimated based on the stellar flux and the distance
of a planet from its host star.  Here we investigate a second-order
effect over $\sim$decade timescales: the {\it change} in insolation as
an M dwarf varies may be crucial for planetary environmental impact.
We find that $\sim$90\% of M dwarfs vary by less than 2\% at
wavelengths where they emit much of their light, implying that
long-term flux levels do not change appreciably on most planets
orbiting red dwarfs.

To further refine the fractions of variables in the red dwarf
population, future studies should target larger, volume-limited
samples.  This would reduce concerns related to biases in the current
sample, including the overabundance of active young stars and
unresolved multiples that have been treated as single stars.  In
addition, more refined studies could probe to higher precision.  For
example, the CTIO/SMARTS 0.9m telescope used to make the observations
discussed here is able to reach photometric precisions of $\sim$1 mmag
when target stars are defocused to boost the total stellar signals,
compared to the 7 mmag variability floor achieved here using the
tightly focused images required for astrometry.
Nonetheless, we now know that most red dwarfs, which comprise
three-quarters of all stars in the solar neighborhood, are
photometrically stable on decade timescales at a level of 2\% at
optical wavelengths.  This bodes well for the stability of
environments on any planets that might be orbiting them.

\section{Acknowledgments}

ADH thanks the McNair Fellowship program under the support of TRIO for
providing funding for this research.  The RECONS effort has been
supported by the National Science Foundation through grants AST
05-07711, AST 09-08402, and AST 14-12026.  We also thank the members
of the SMARTS Consortium and the CTIO staff, who enable the operations
of the small telescopes at CTIO.  This research has made use of
results from the SAO/NASA Astrophysics Data System Bibliographic
Services, as well as the SIMBAD and VizieR databases operated at CDS,
Strasbourg, France, and the Two Micron All Sky Survey, which is a
joint project of the University of Massachusetts and the Infrared
Processing and Analysis Center, funded by NASA and NSF.

\clearpage


\clearpage


\pagestyle{empty}


\hoffset000pt{}
\voffset100pt{}
\centering
\begin{deluxetable}{clcccrrrcrrcrcrrrrl}
\rotate
\tablewidth{0pt}
\tabletypesize{\tiny}
\tablecaption{Red Dwarfs Studied for Long-Term Variability}
\tablehead{\vspace{-10pt} \\
\multicolumn{1}{c}{}&
\multicolumn{1}{c}{Name}&
\multicolumn{1}{c}{Young}&
\multicolumn{1}{c}{R.A.} &
\multicolumn{1}{c}{Dec.} &
\multicolumn{1}{c}{$V$} &
\multicolumn{1}{c}{$R$} &
\multicolumn{1}{c}{$I$} &
\multicolumn{1}{c}{Ref.} &
\multicolumn{1}{c}{$K_s$} &
\multicolumn{1}{c}{$V-K_s$} &
\multicolumn{1}{c}{Filter} &
\multicolumn{1}{c}{Variability} &
\multicolumn{1}{c}{Timespan} &
\multicolumn{1}{c}{Duration} &
\multicolumn{1}{c}{$\#$Nights} &
\multicolumn{1}{c}{$\#$Frames} &
\multicolumn{1}{c}{$\pi_{trig}$ $\pm$ $\sigma$} &
\multicolumn{1}{c}{Ref.} \\
\colhead{} &
\colhead{} &
\colhead{} &
\colhead{J2000.0} &
\colhead{J2000.0} &
\colhead{} &
\colhead{} &
\colhead{} &
\colhead{} &
\colhead{} &
\colhead{} &
\colhead{} &
\colhead{(mmag)} &
\colhead{} &
\colhead{(yr)} &
\colhead{} &
\colhead{} &
\colhead{(mas)} &
\colhead{} \\
\colhead{} &
\colhead{(1)} &
\colhead{(2)} &
\colhead{(3)} &
\colhead{(4)} &
\colhead{(5)} &
\colhead{(6)} &
\colhead{(7)} &
\colhead{(8)} &
\colhead{(9)} &
\colhead{(10)} &
\colhead{(11)} &
\colhead{(12)} &
\colhead{(13)} &
\colhead{(14)} &
\colhead{(15)} &
\colhead{(16)} &
\colhead{(17)} &
\colhead{(18)}}
\startdata
........  & GJ 1001A         &...& 00 04 36  & $-$40 44 02 &  12.83 & 11.62 & 11.08 &  Win15  &  7.74 &  5.09 &  R  &  13.7  &  1999.64-2011.74  &  12.10  &  21  & 112  &  77.02  2.07  &  Die14  \\ %
          & GJ 1002          &...& 00 06 43  & $-$07 32 17 &  13.84 & 12.21 & 10.21 &  Win15  &  7.44 &  6.40 &  R  &  13.9  &  2003.77-2012.87  &   9.10  &  14  &  64  & 207.18  3.09  &  Dav14  \\ %
          & L 217-028        &...& 00 08 17  & $-$57 05 52 &  12.13 & 11.00 &  9.57 &  Win15  &  7.40 &  4.73 &  V  &   9.8  &  2000.50-2008.00  &   7.25  &  15  &  68  &       ...     &   ...   \\ 
          & G 131-026        &...& 00 08 53  & $+$20 50 25 &  13.52 & 12.19 & 10.50 &  Rie14  &  8.01 &  5.51 &  V  &  18.9  &  1999.71-2011.89  &  12.18  &  21  &  97  &  54.13  1.35  &  Rie14  \\ %
          & GJ 1005AB        &...& 00 15 28  & $-$16 08 01 &  11.48 & 10.27 &  8.70 &  Win15  &  6.39 &  5.09 &  V  &   8.2  &  2003.75-2011.90  &   7.96  &  22  &  95  &       ...     &   ...   \\ 
          & GJ 2006A         & Y & 00 27 50  & $-$32 33 06 &  12.95 & 11.79 & 10.29 &  Rie14  &  8.01 &  4.94 &  V  &  76.8  &  2000.57-2010.82  &  10.25  &  17  &  69  &  30.14  2.50  &  Rie14  \\ %
          & GJ 2006B         & Y & 00 27 50  & $-$32 33 23 &  13.25 & 12.04 & 10.48 &  Rie14  &  8.12 &  5.13 &  V  &  36.2  &  2000.57-2010.82  &  10.25  &  17  &  69  &  31.78  2.47  &  Rie14  \\ %
          & GR 50            &...& 00 32 53  & $-$04 34 07 &  13.97 & 12.68 & 11.00 &  Win15  &  8.35 &  5.62 &  R  &  13.0  &  2007.82-2012.95  &   5.13  &  11  &  59  &       ...     &   ...   \\
\enddata
\vskip10pt

\scriptsize {Notes: ``Y'' in Column 2 indicates stars confirmed to be
young in \citet{2014AJ....147...85R}.  $VRI$ photometry is from this
paper = *, \citet{2014AJ....147...94D} = Die14,
\citet{2004AJ....128.2460H} = Hen04, \citet{2006AJ....132.2360H} =
Hen06, \citet{2005AJ....129.1954J} = Jao05,
\citet{2011AJ....141..117J} = Jao11, \citet{2014AJ....147..21J} =
Jao14, \citet{2013AJ....146..154M} = Mam13,
\citet{2010AJ....140..897R} = Rie10, \citet{2014AJ....147...85R} =
Rie14, \citet{2009AJ....137.4547S} = Sub09,
\citet{2011AJ....141...21W} = Win11, and 
\citet{2015AJ....149....5W} = Win15.}


\end{deluxetable}

\clearpage

\hoffset000pt{}
\voffset000pt{}
\begin{deluxetable}{lcrrrccccccc}
\tabletypesize{\tiny}
\tablecaption{M Dwarfs Nearer than 25 Parsecs with Exoplanets}
\tablewidth{0pt}
\tablehead{\vspace{-8pt} \\
\multicolumn{1}{c}{Name}&
\multicolumn{1}{c}{\# Planets}&
\multicolumn{1}{c}{$V$} &
\multicolumn{1}{c}{$R$} &
\multicolumn{1}{c}{$I$} &
\multicolumn{1}{c}{Ref.} &
\multicolumn{1}{c}{$V-K_s$} &
\multicolumn{1}{c}{Filter} &
\multicolumn{1}{c}{Variability} &
\multicolumn{1}{c}{Duration} &
\multicolumn{1}{c}{$\#$ Nights} &
\multicolumn{1}{c}{$\#$ Frames} \\
\colhead{           } &
\colhead{           } &
\colhead{           } &
\colhead{           } &
\colhead{           } &
\colhead{           } &
\colhead{           } &
\colhead{           } &
\colhead{(mmag)     } &
\colhead{(yr)       } &
\colhead{           } &
\colhead{           }}

\startdata
 
GJ 163      & 3 & 11.84 & 10.76 &  9.34 & Bes90 &  4.71 & ... &   ...  &  ...   & ...  &   ... \\  
            &   &       &       &       &       &       &     &        &        &      &       \\  %
GJ 176      & 1 &  9.98 &  8.95 &  7.72 & Wei96 &  4.37 & ... &   ...  &  ...   & ...  &   ... \\  
            &   &       &       &       &       &       &     &        &        &      &       \\  %
GJ 179      & 1 & 11.98 & 10.83 &  9.33 & Bes90 &  5.04 & ... &   ...  &  ...   & ...  &   ... \\  
            &   &       &       &       &       &       &     &        &        &      &       \\  %
GJ 317      & 1 & 12.01 & 10.84 &  9.37 & Win15 &  4.98 &  R  &  17.4  &  4.35  &  15  &   75  \\  
            &   &       &       &       &       &       &     &        &        &      &       \\
\enddata
\vskip10pt

\scriptsize {Notes: $VRI$ photometry is from this paper = *,
  \citet{1990A&AS...83..357B} = Bes90, \citet{2010MNRAS.403.1949K} =
  Koe10, \citet{1996AJ....112.2300W} = Wei96, and our program,
  \citet{2010AJ....140..897R} = Rie10, \citet{2015AJ....149....5W} =
  Win15.  $V-K_s$ uses $K_s$ from 2MASS.  For seven stars, the filter
  listed is used to observe the star for parallax and photometric
  variability.  The two stars indicated with $^a$ are being observed
  but not included in the larger study discussed here because they do
  not yet have four years of observations.  Stars not observed in this
  program are listed for completeness, with the filter, variability,
  and coverage metrics left blank.}

\end{deluxetable}

\clearpage


\begin{figure}
\label{fig.hist.coverage}
\vspace{-1.0in}
\hspace{-1.0in}
\includegraphics[angle=270,scale=0.80]{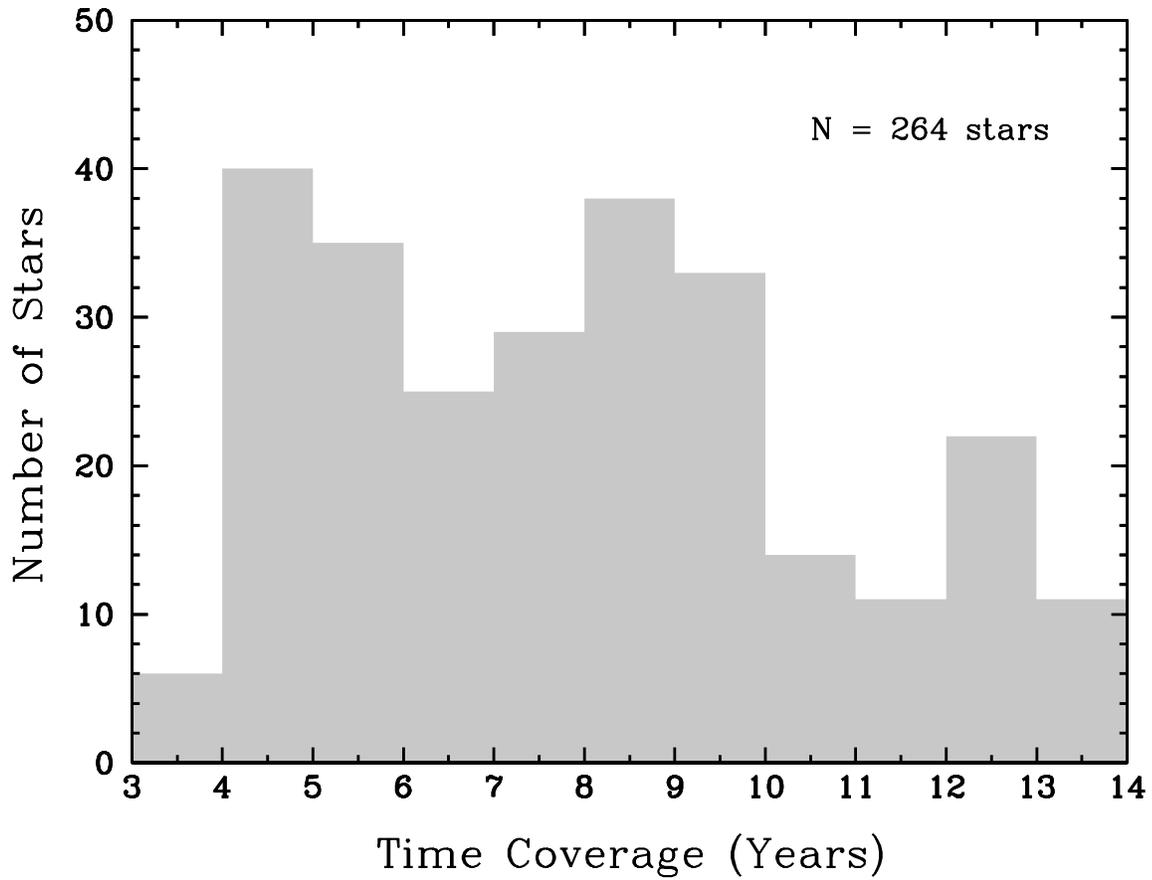}

\caption{This histogram outlines the photometric variability time
  coverage for the sample stars, listed individually in Table 1.  The
  median time coverage is 7.9 years.  Some stars were not observed
  uniformly throughout the time period, i.e.~they had fewer than three
  nights of observations in some years.}

\end{figure}

 
\begin{figure}
\label{fig.varVRI.ps}
\vspace{-1.0in}
\hspace{-0.2in}
\includegraphics[angle=0,scale=0.80]{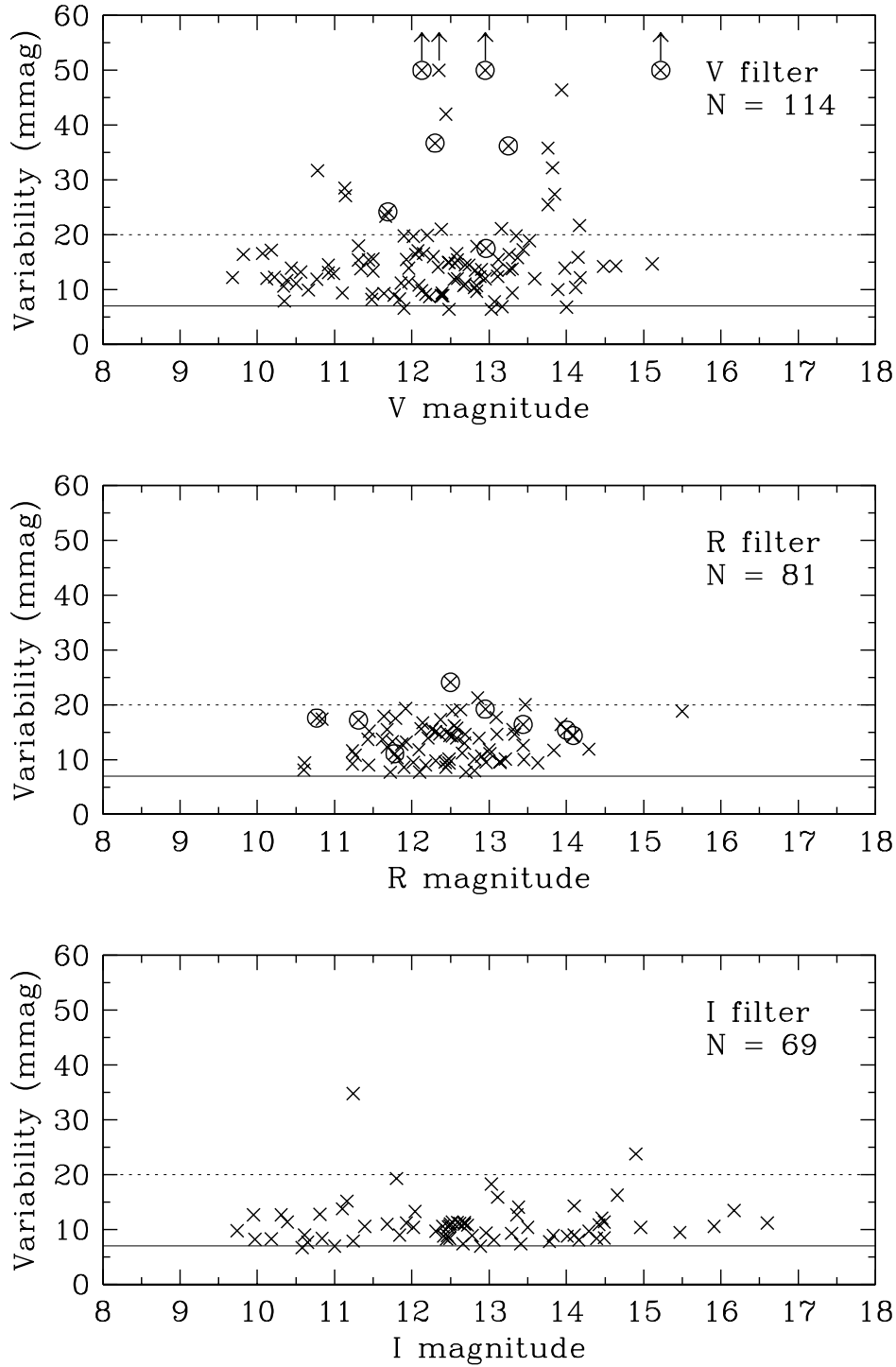}

\vspace{-0.3in}

\caption{The photometric variability defined as the standard
deviation of a target star's brightness relative to reference stars
in the field, is shown in each of the three filters used for
long-term astrometric measurements, $VRI$, as a function of target
brightness. Circled points represent stars denoted as young, with
``Y'' beside their names in Table 1.  Four stars discussed in the
text vary by more than 50 mmag at $V$ and are represented by arrowed
points in the $V$ panel.  There are no obvious trends in photometric
variability with apparent magnitude, with a variability floor
represented by solid lines at 7 mmag in all three filters.  Dotted
lines at 20 mmag indicate our selected threshold between active
(above) and quiescent (below) stars.}

\end{figure}


\begin{figure}
\label{fig.varV-K.ps}
\vspace{-1.0in}
\hspace{-0.2in}
\includegraphics[angle=0,scale=0.80]{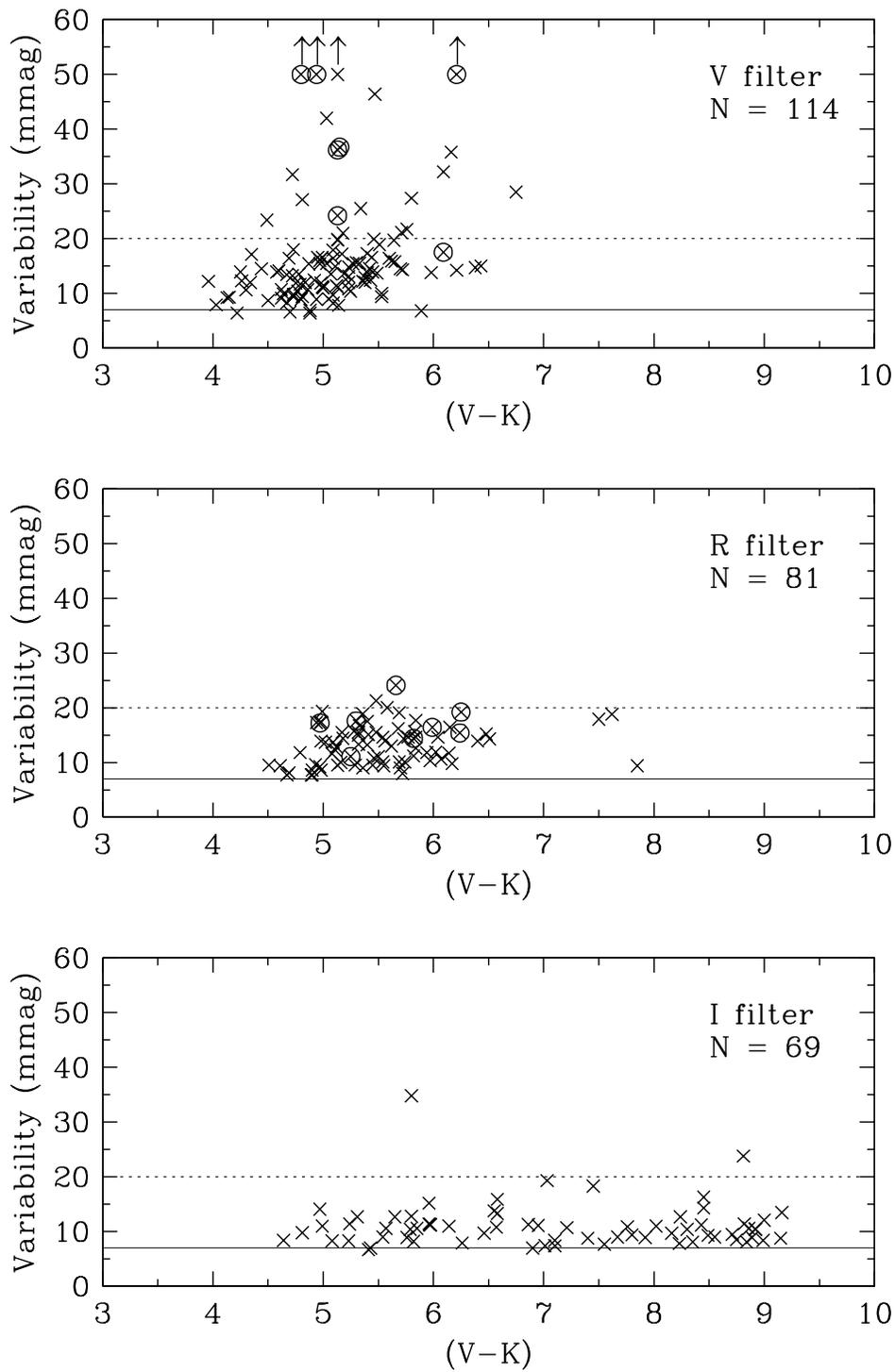}

\vspace{-0.3in}

\caption{The photometric variability is shown in each of the three
  filters used for long-term astrometric measurements, $VRI$, as a
  function of target $V-K$ color.  There are no obvious trends in
  photometric variability with temperature.  See the caption of Figure
  2 for additional details of the plot format.}

\end{figure}


\begin{figure}
\label{fig.varMv.ps}
\vspace{-1.0in}
\hspace{-0.2in}
\includegraphics[angle=0,scale=0.80]{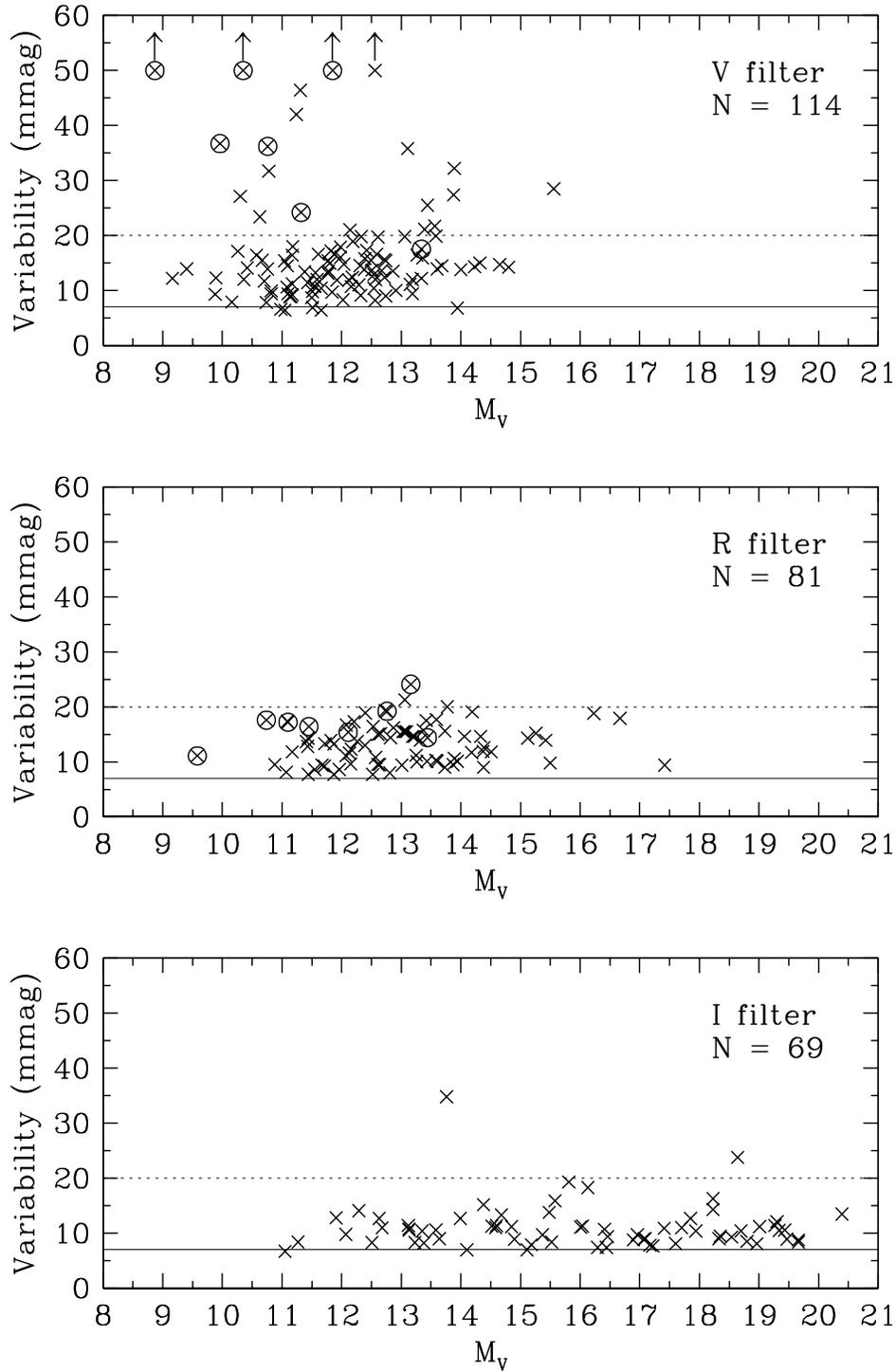}

\caption{The photometric variability is shown in each of the three
  filters used for long-term astrometric measurements, $VRI$, as a
  function of absolute magnitude.  There are no obvious trends in
  photometric variability with luminosity.  See the caption of Figure
  2 for additional details of the plot format.}

\end{figure}


\begin{figure}
\label{fig.hist.varMv}
\vspace{-1.0in}
\hspace{-1.0in}
\includegraphics[angle=270,scale=0.80]{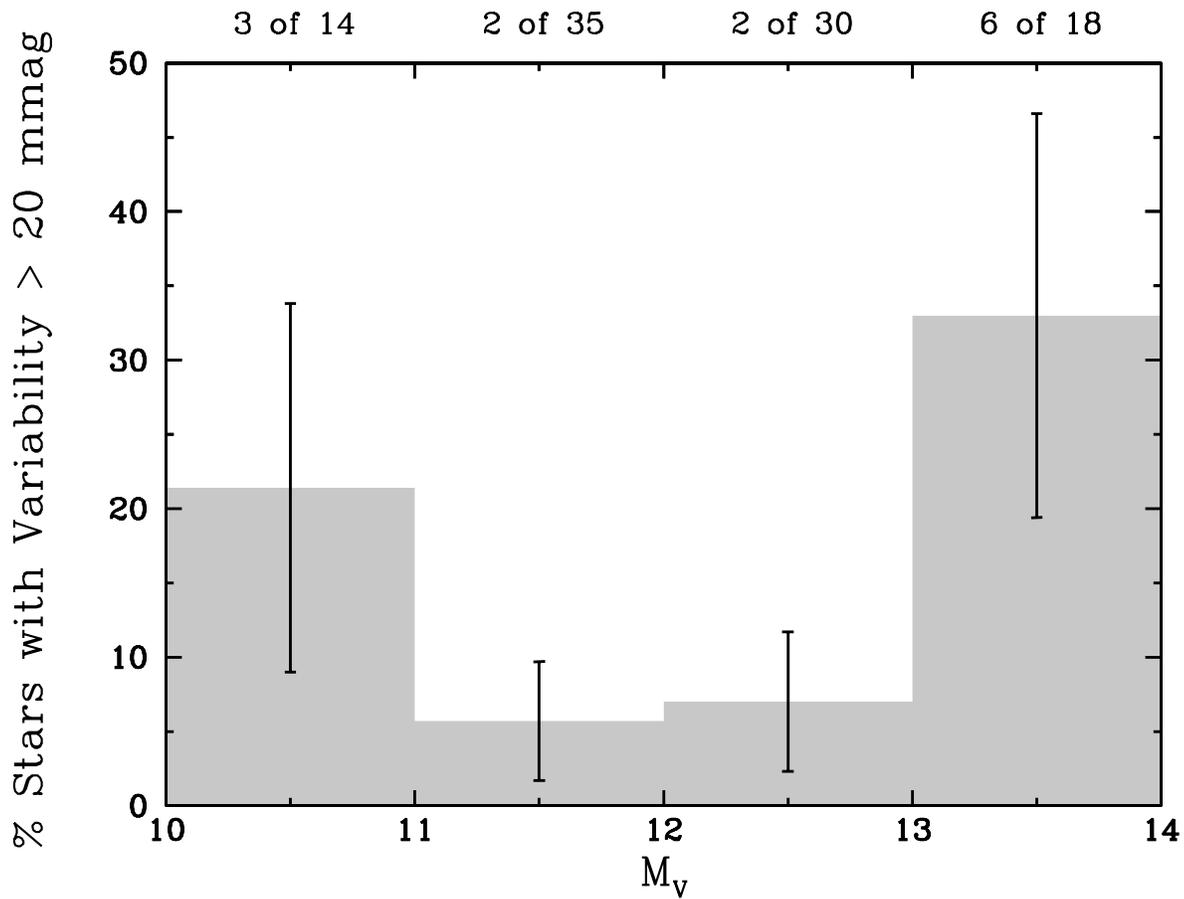}

\caption{The fractions of variable stars within 25 pc observed in the
  $V$ filter are shown, setting the variability threshold at 20 mmag.
  In total, there are 106 stars in the sample, of which 14 (13\%) vary
  by more than 20 mmag.  Nine stars are not shown because they fall in
  one brighter bin and two fainter bins with too few stars to be
  statistically useful.  Large error bars represent counting
  statistics, illustrating that the sample suffers from small numbers
  of detected variable stars.  This precludes detecting any clear
  trends in variability in the current dataset.}

\end{figure}


\begin{figure}[ht]
\vskip-10pt
\begin{center}

{\includegraphics[scale=0.28,angle=270]{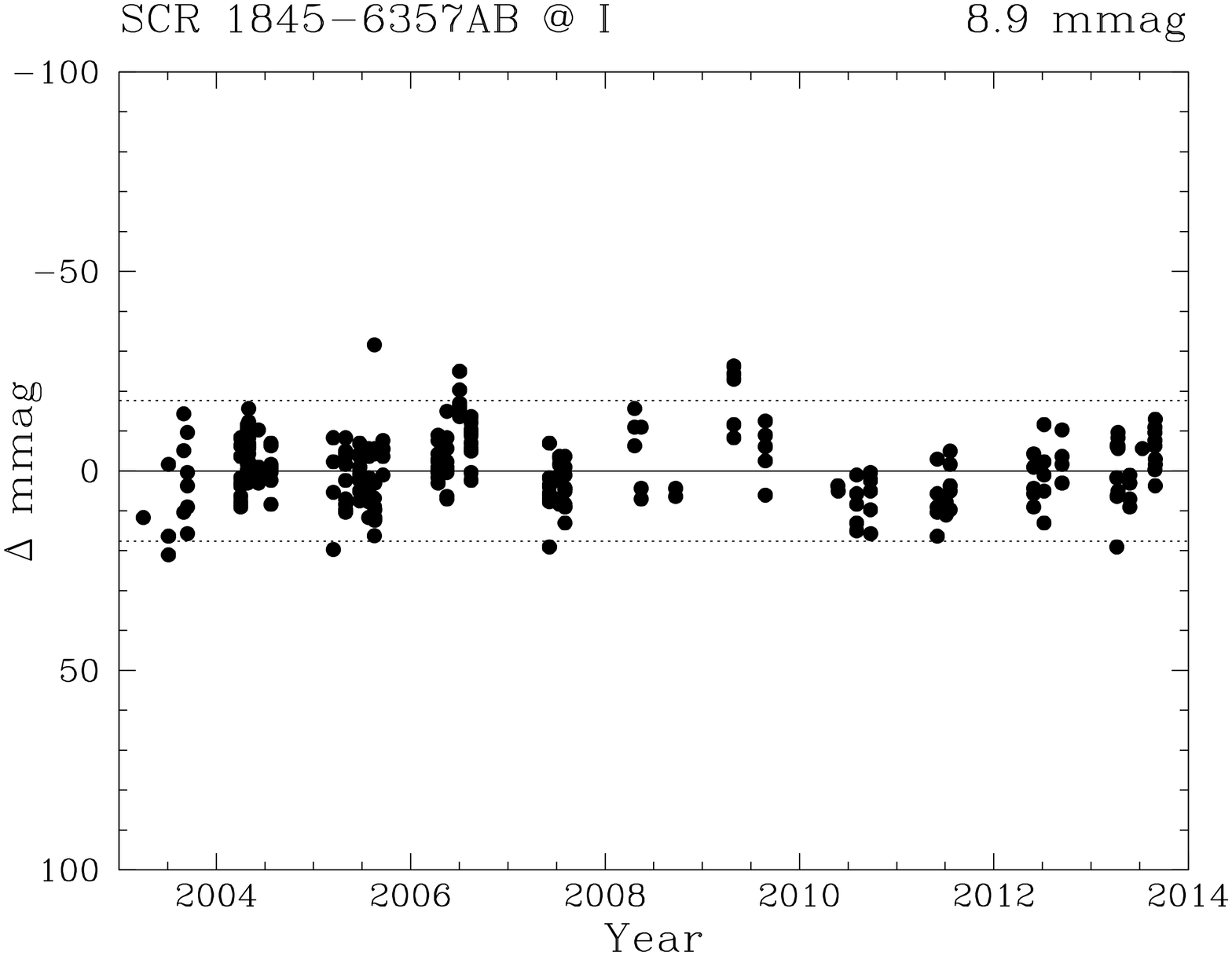}}
\hskip15pt
{\includegraphics[scale=0.28,angle=270]{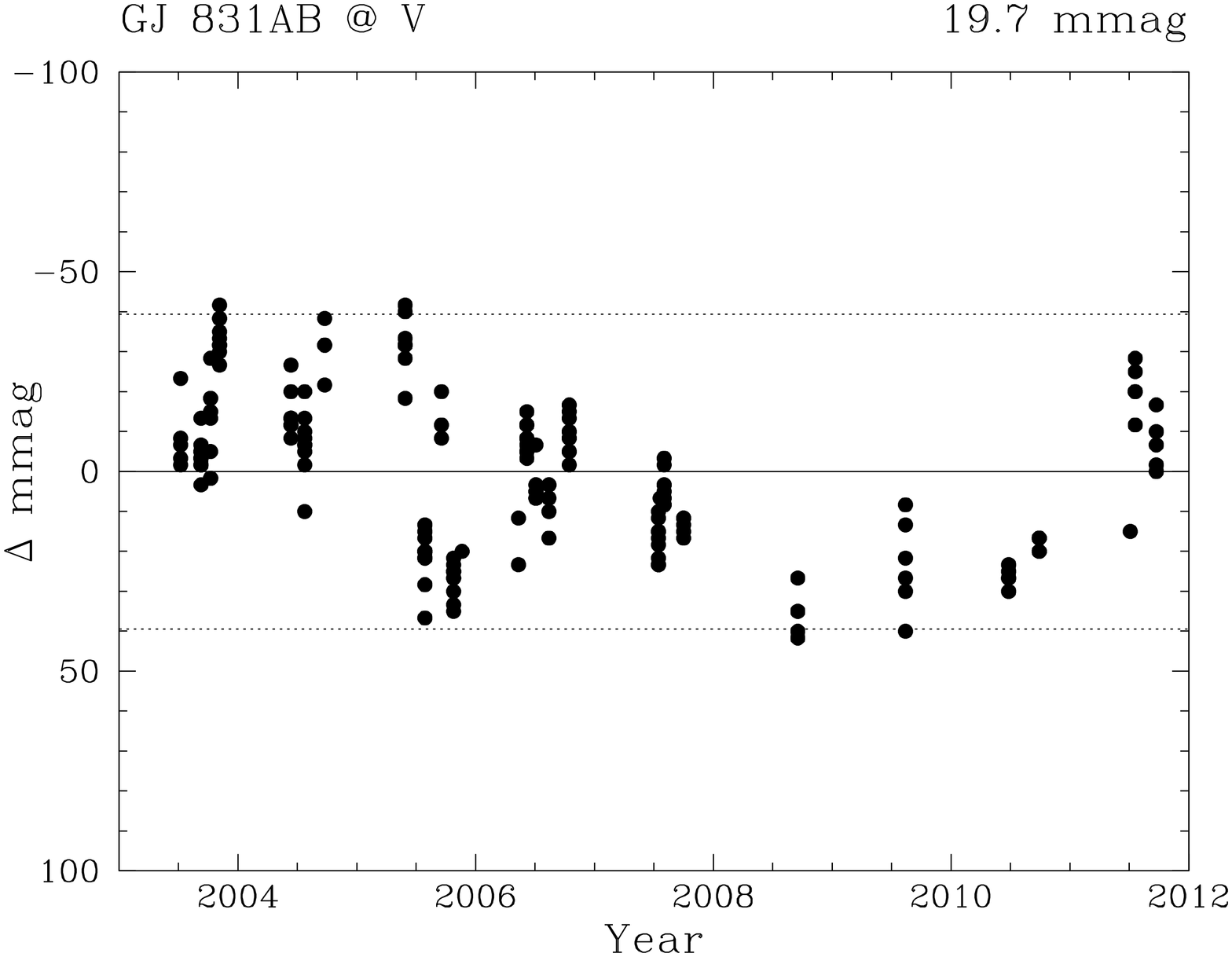}}
\hskip-10pt
{\includegraphics[scale=0.28,angle=270]{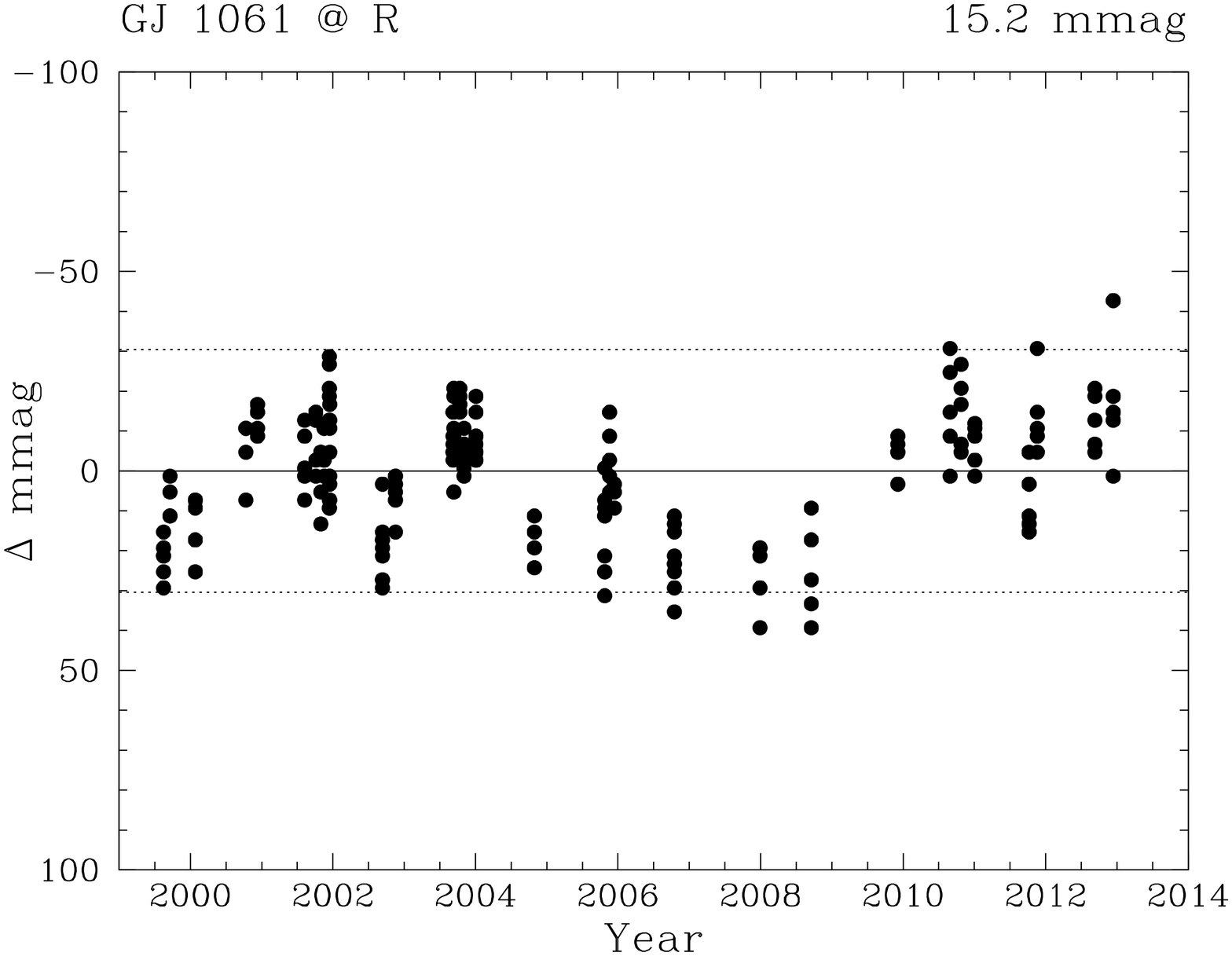}}
\hskip15pt
{\includegraphics[scale=0.28,angle=270]{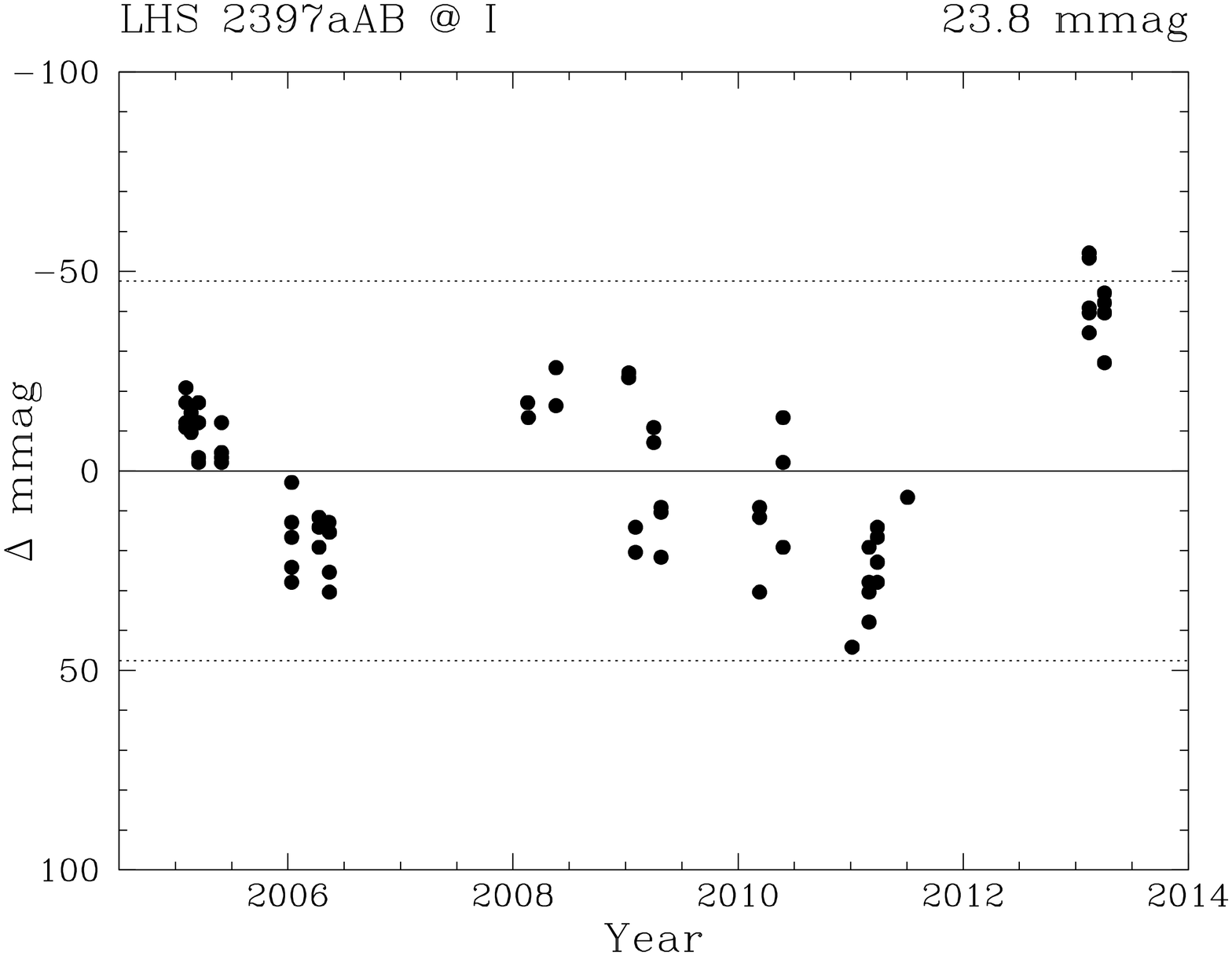}}
\hskip-10pt
{\includegraphics[scale=0.28,angle=270]{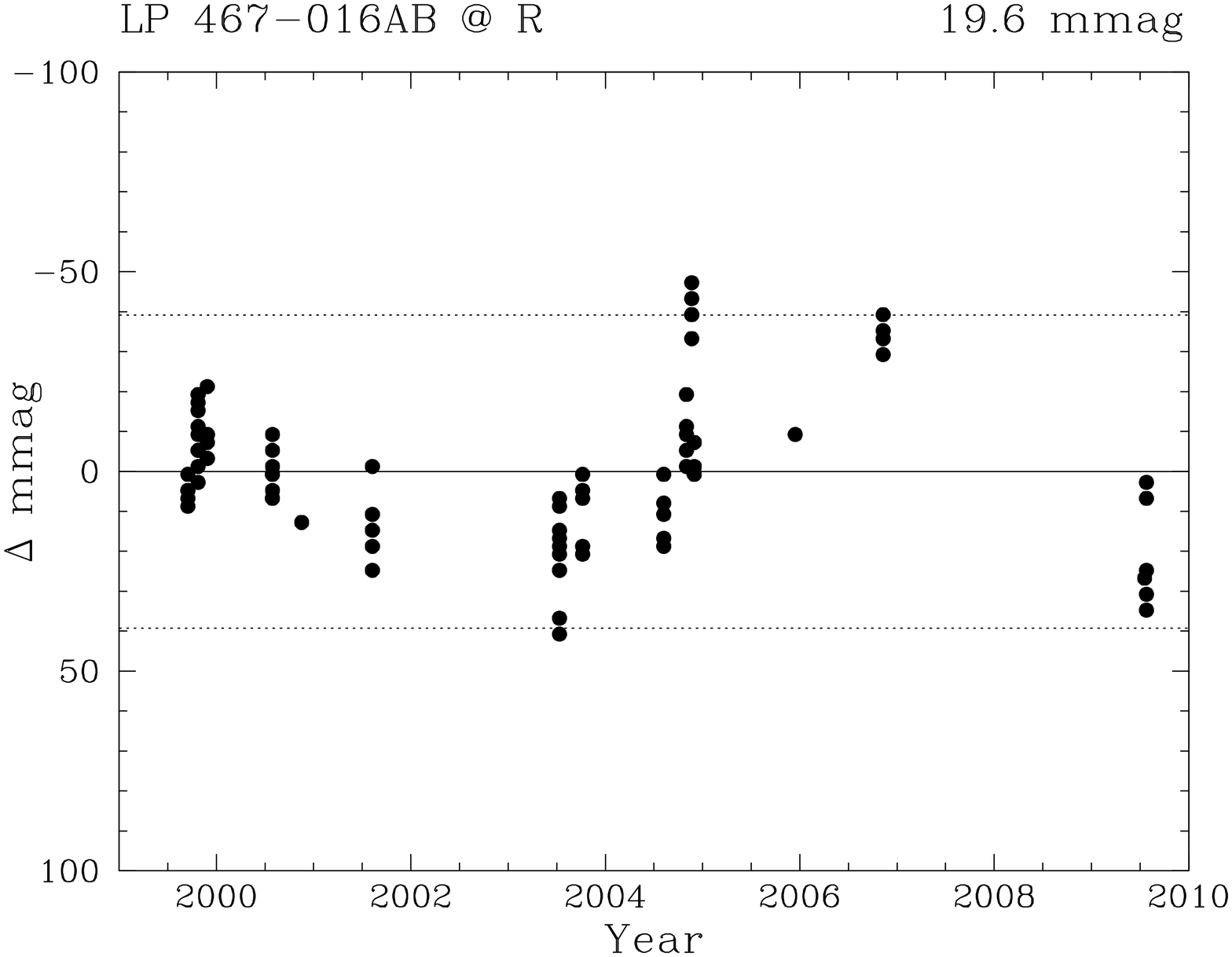}}
\hskip15pt
{\includegraphics[scale=0.28,angle=270]{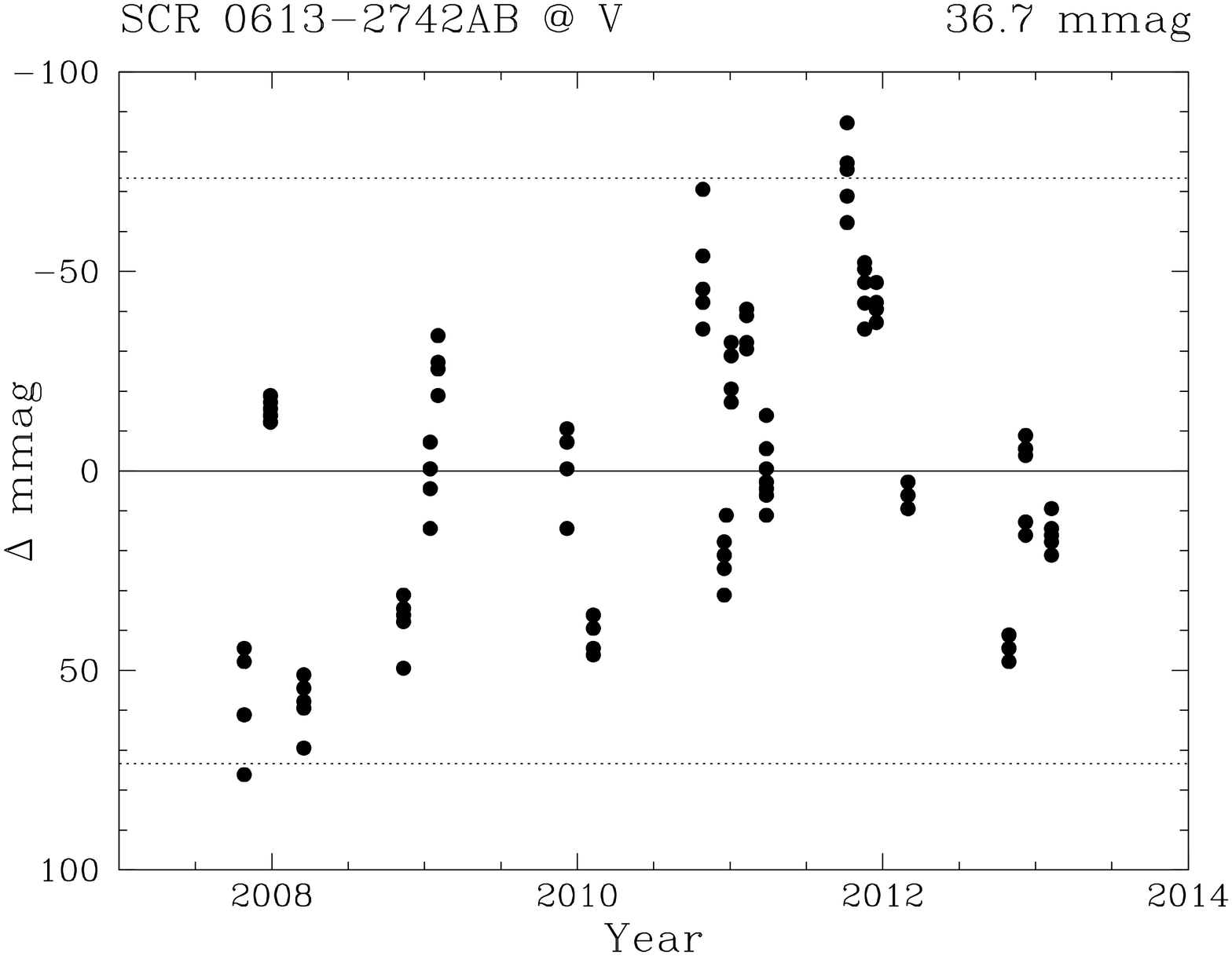}}

\vskip20pt

\caption{Each panel illustrates the brightness for one target
  measured relative to a set of observed background reference stars,
  where each point represents an individual image.  The filter for
  the observations is given in the upper left of each panel after
  the target name.  The relative brightnesses are given as magnitude
  differences, measured in milli-magnitudes, with the average
  deviation from zero listed at the top right.  The star is brighter
  when points are plotted toward the top, i.e. at negative offsets.
  Dotted lines represent offsets twice that of the average
  deviation.  The upper left panel for SCR 1845-6357AB shows a
  non-variable star.  The other five panels show stars with periodic
  variations.}

\end{center}
\end{figure}


\begin{figure}[ht]
\vskip-10pt
\begin{center}

{\includegraphics[scale=0.28,angle=270]{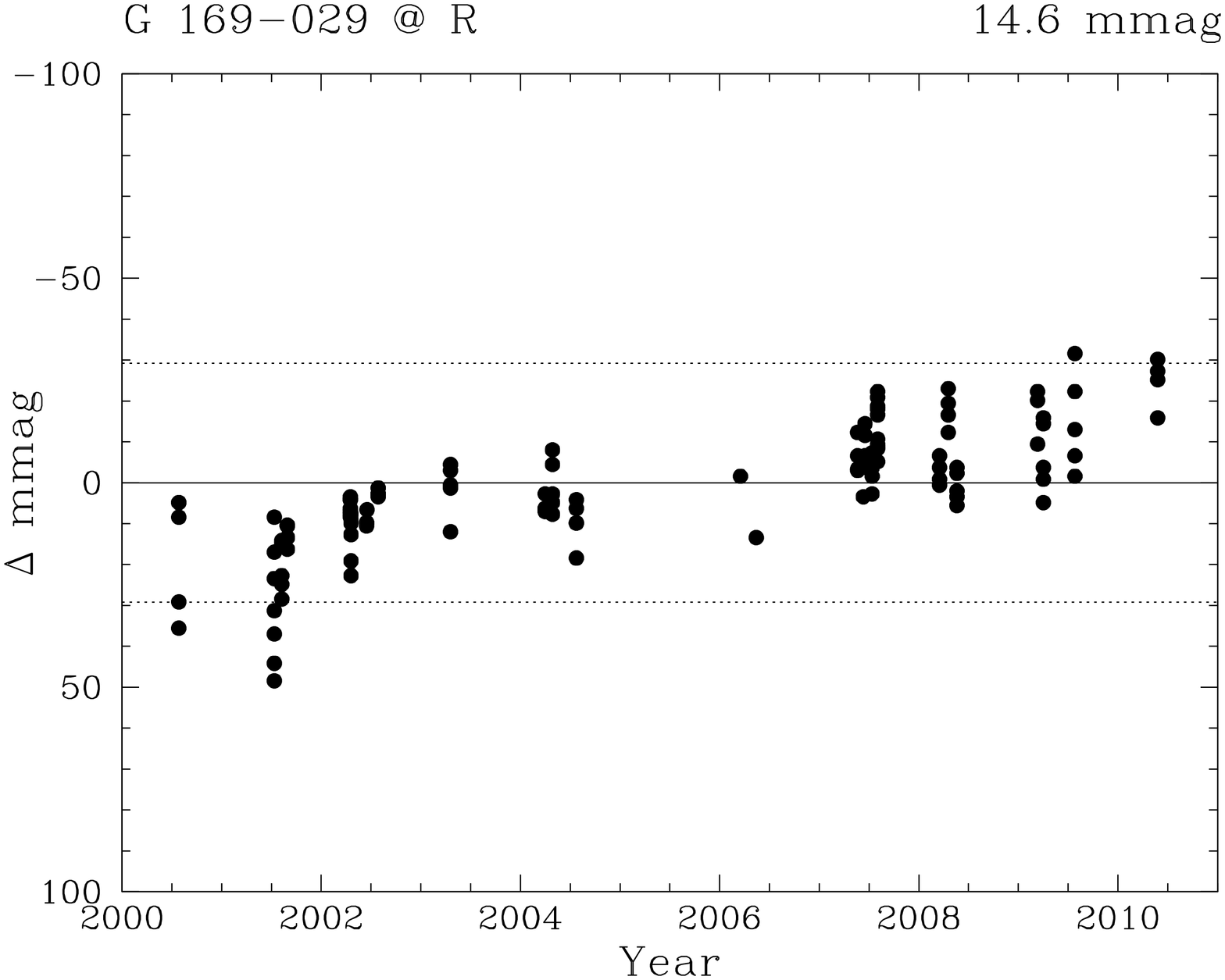}}
\hskip15pt
{\includegraphics[scale=0.28,angle=270]{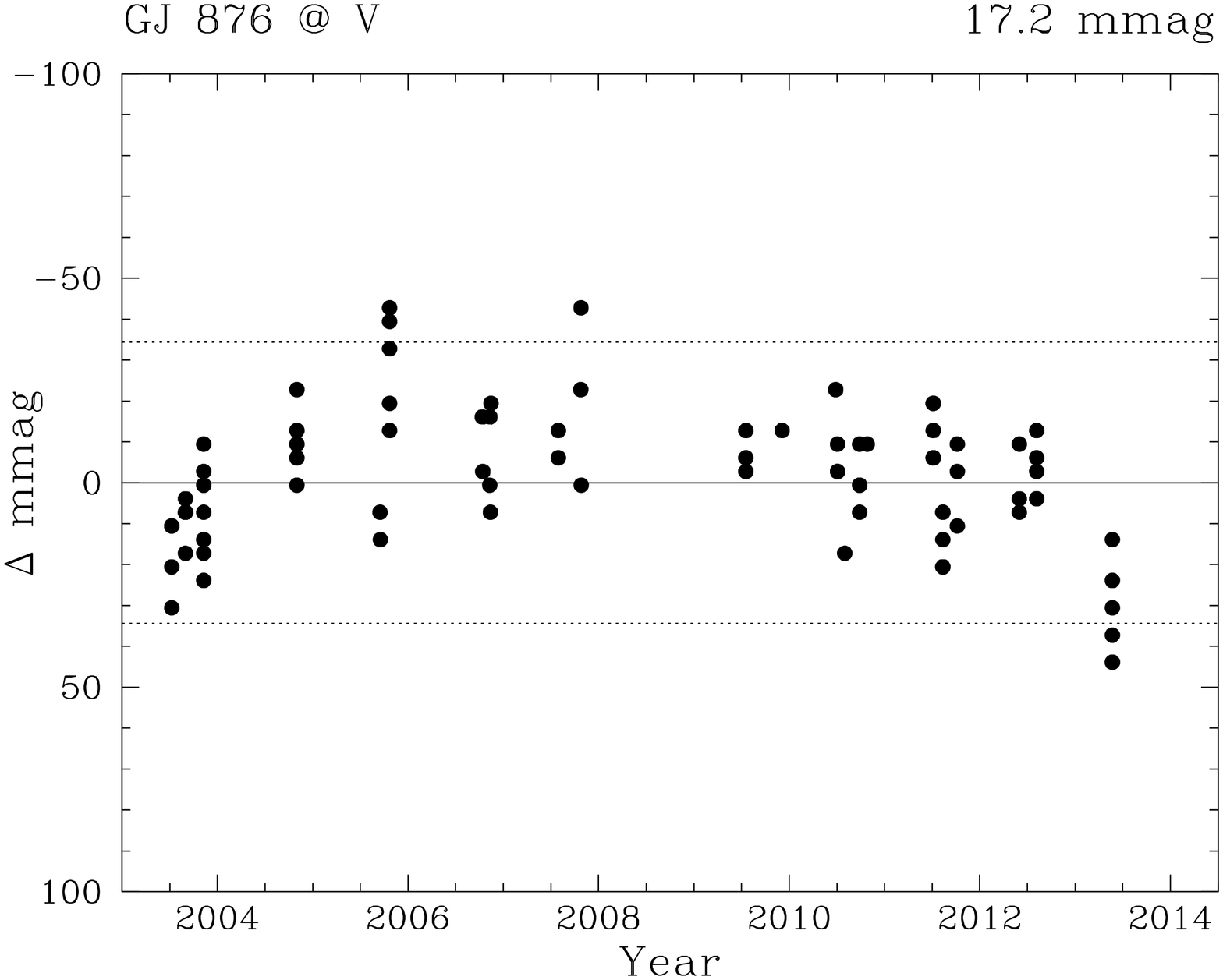}}
\hskip-10pt
{\includegraphics[scale=0.28,angle=270]{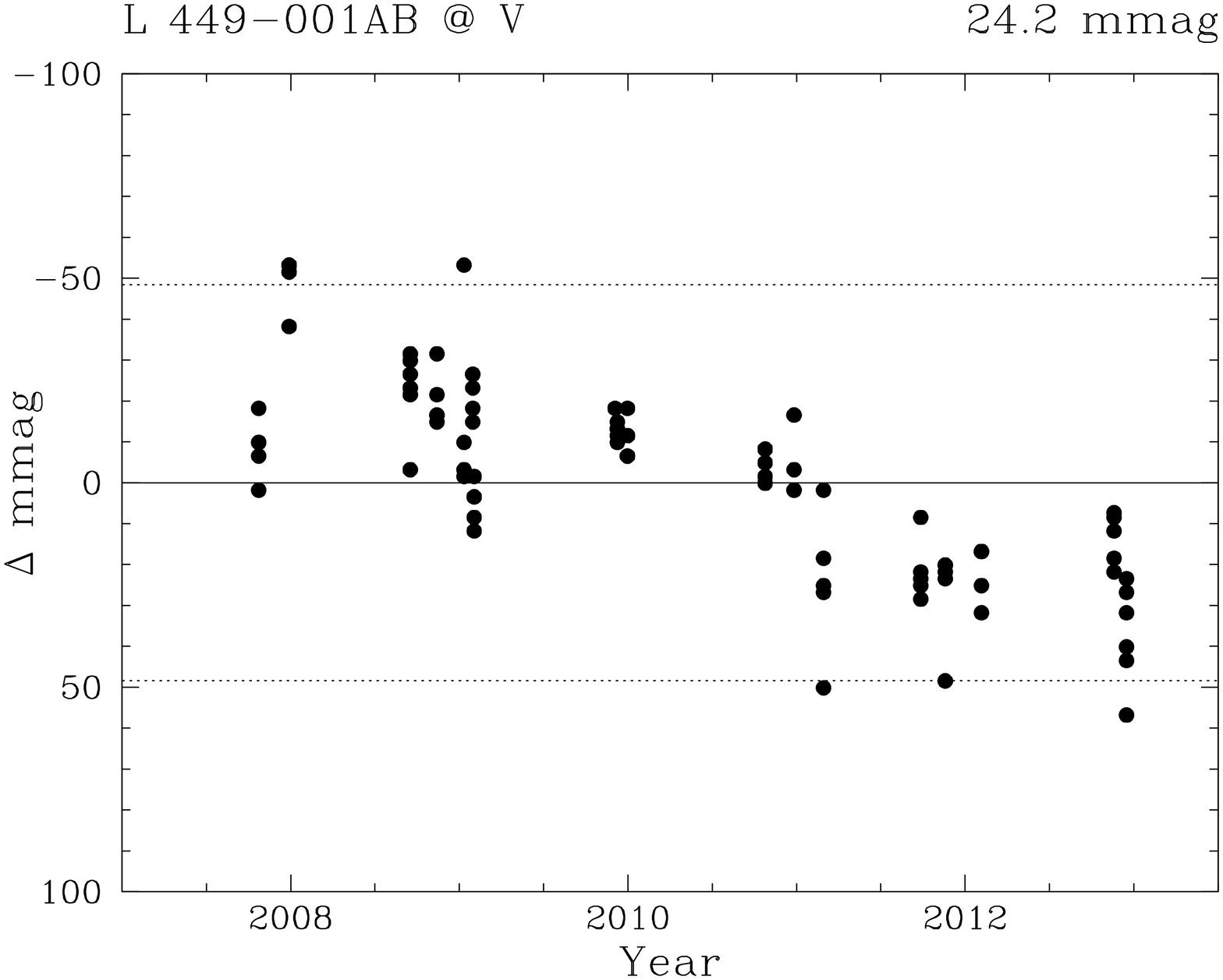}}
\hskip15pt
{\includegraphics[scale=0.28,angle=270]{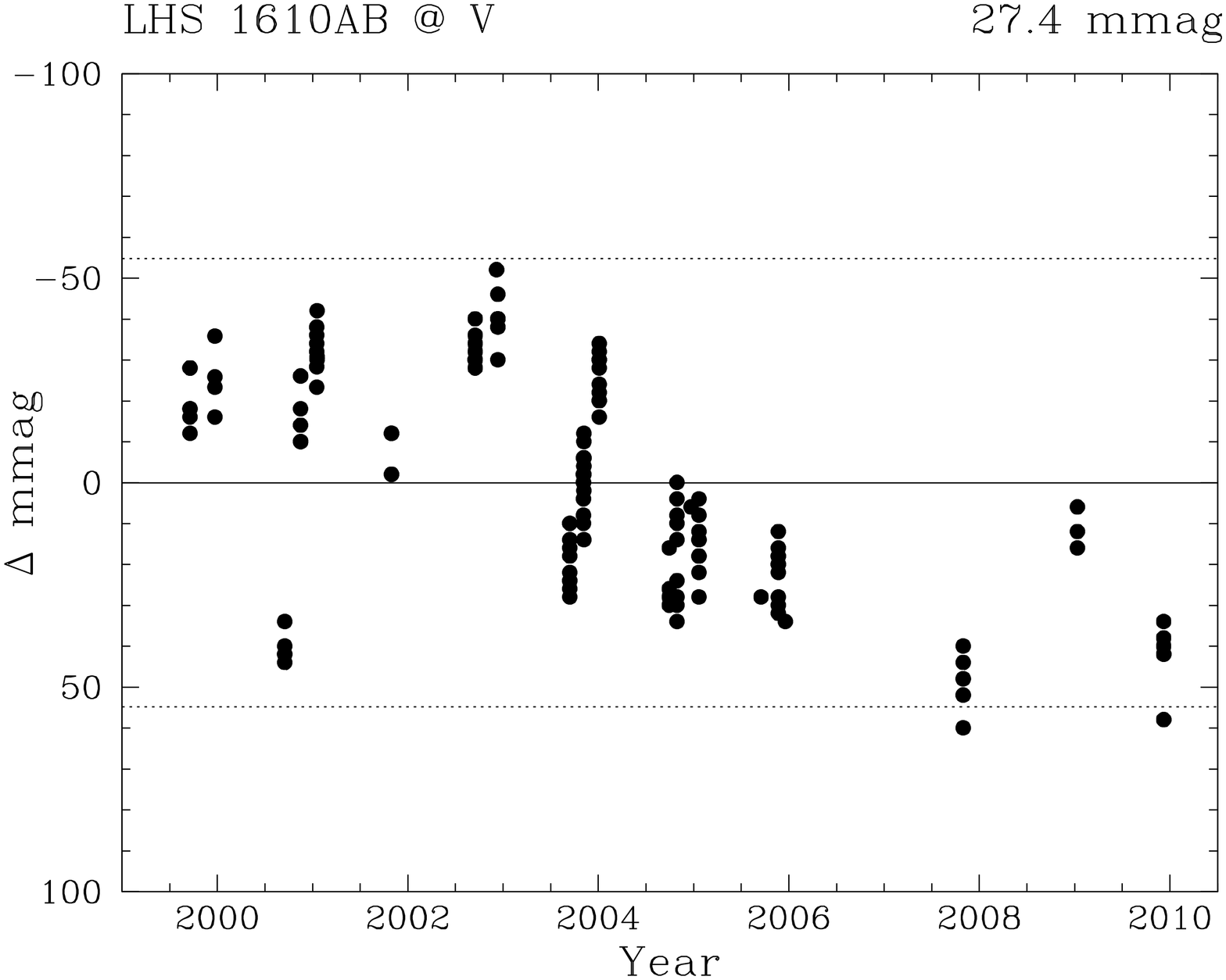}}
\hskip-10pt
{\includegraphics[scale=0.28,angle=270]{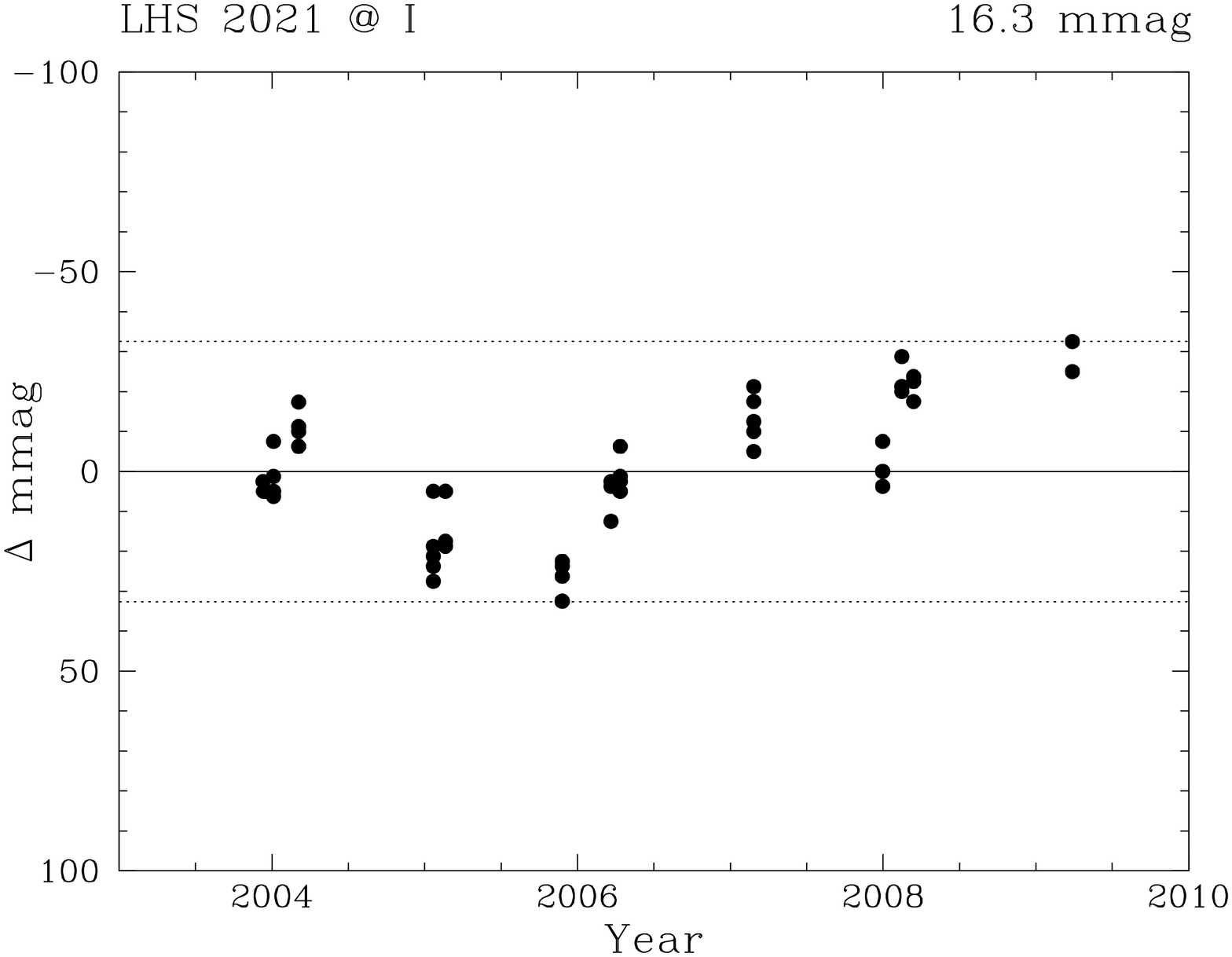}}
\hskip15pt
{\includegraphics[scale=0.28,angle=270]{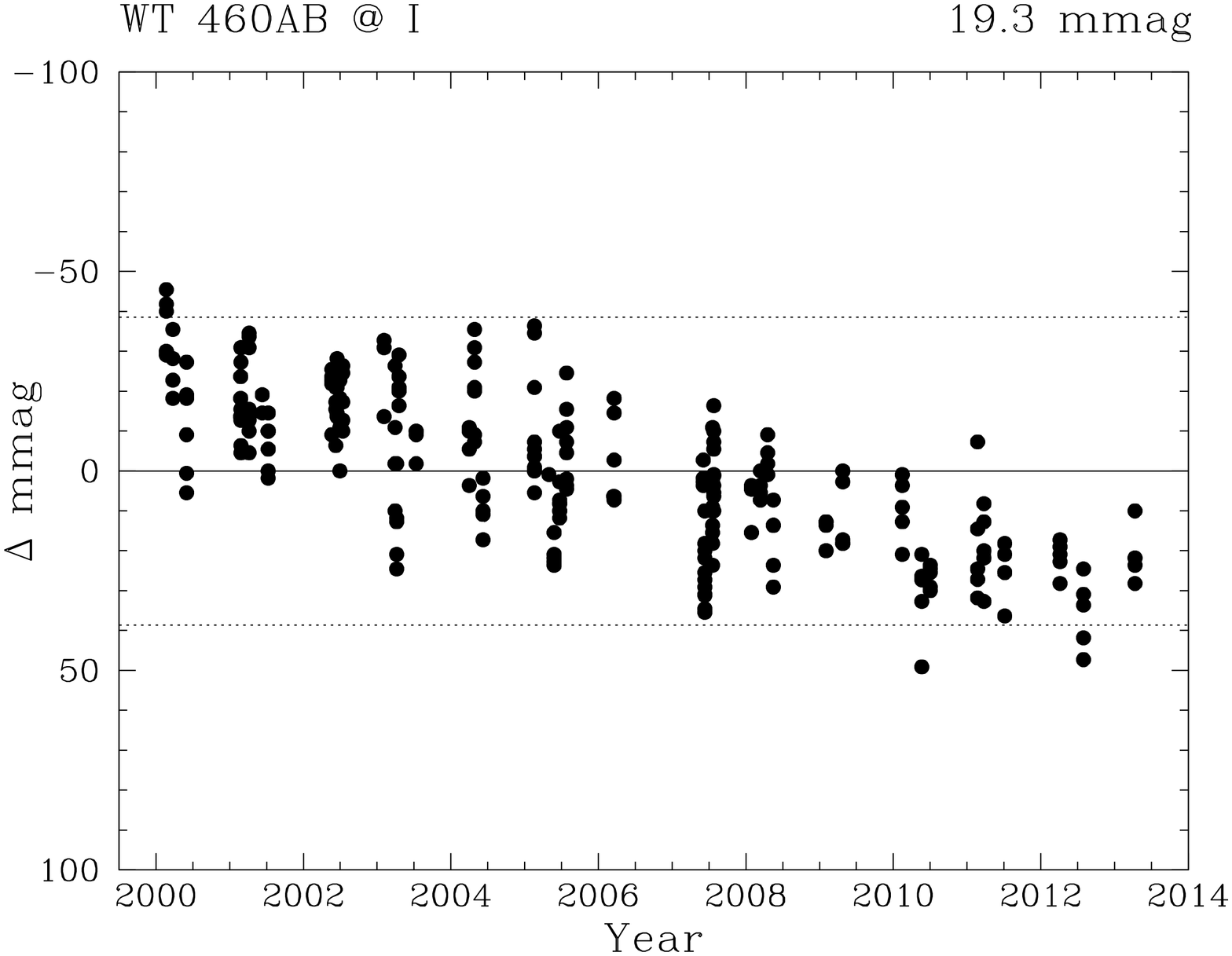}}

\vskip20pt

\caption{See Figure 6 caption for plot layout details. Each of the six
  stars shown exhibits a long-term trend that does not appear to have
  yet completed a full cycle in the available datasets.}

\end{center}
\end{figure}


\begin{figure}[ht]
\vskip-10pt
\begin{center}

{\includegraphics[scale=0.28,angle=270]{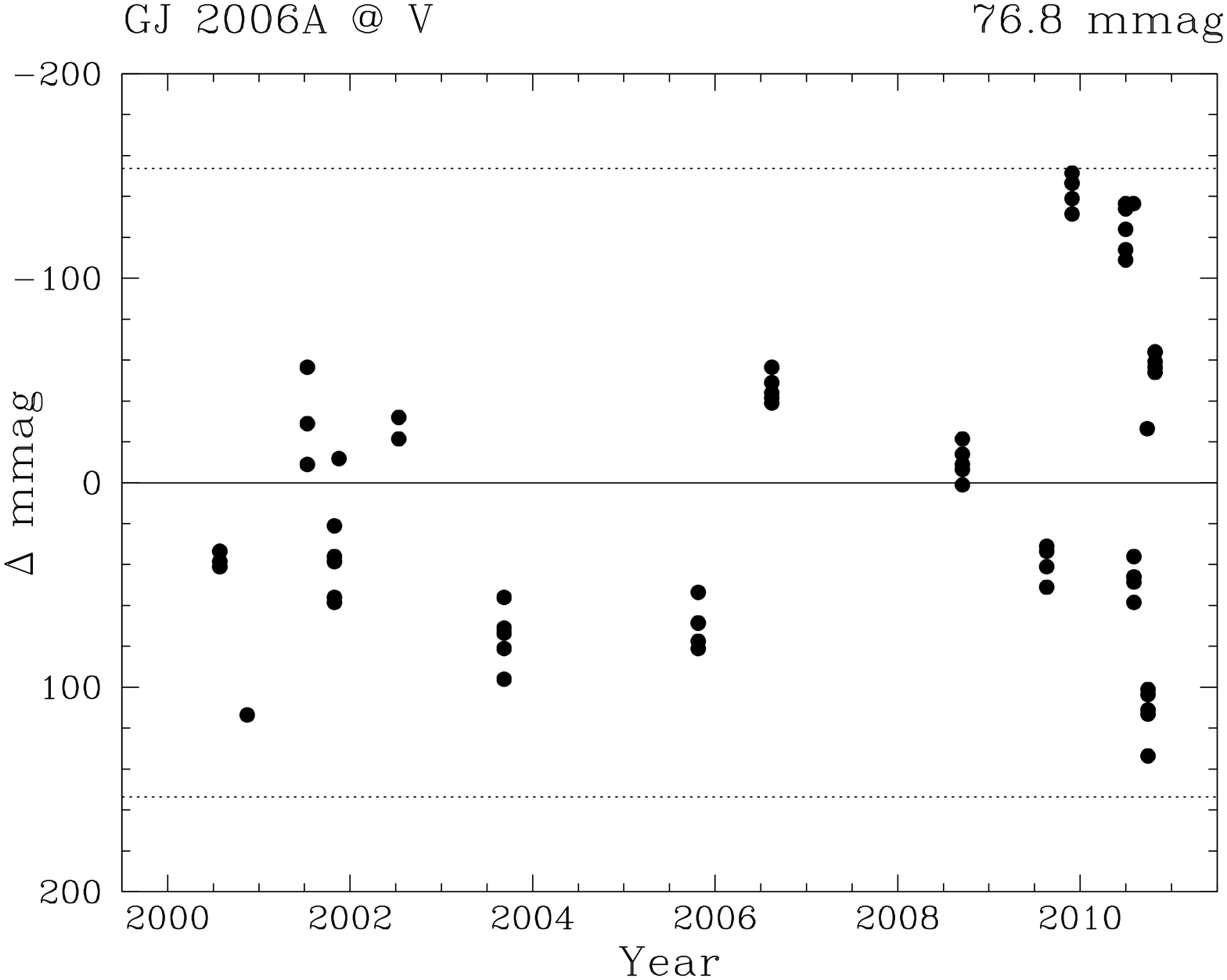}}
\hskip15pt
{\includegraphics[scale=0.28,angle=270]{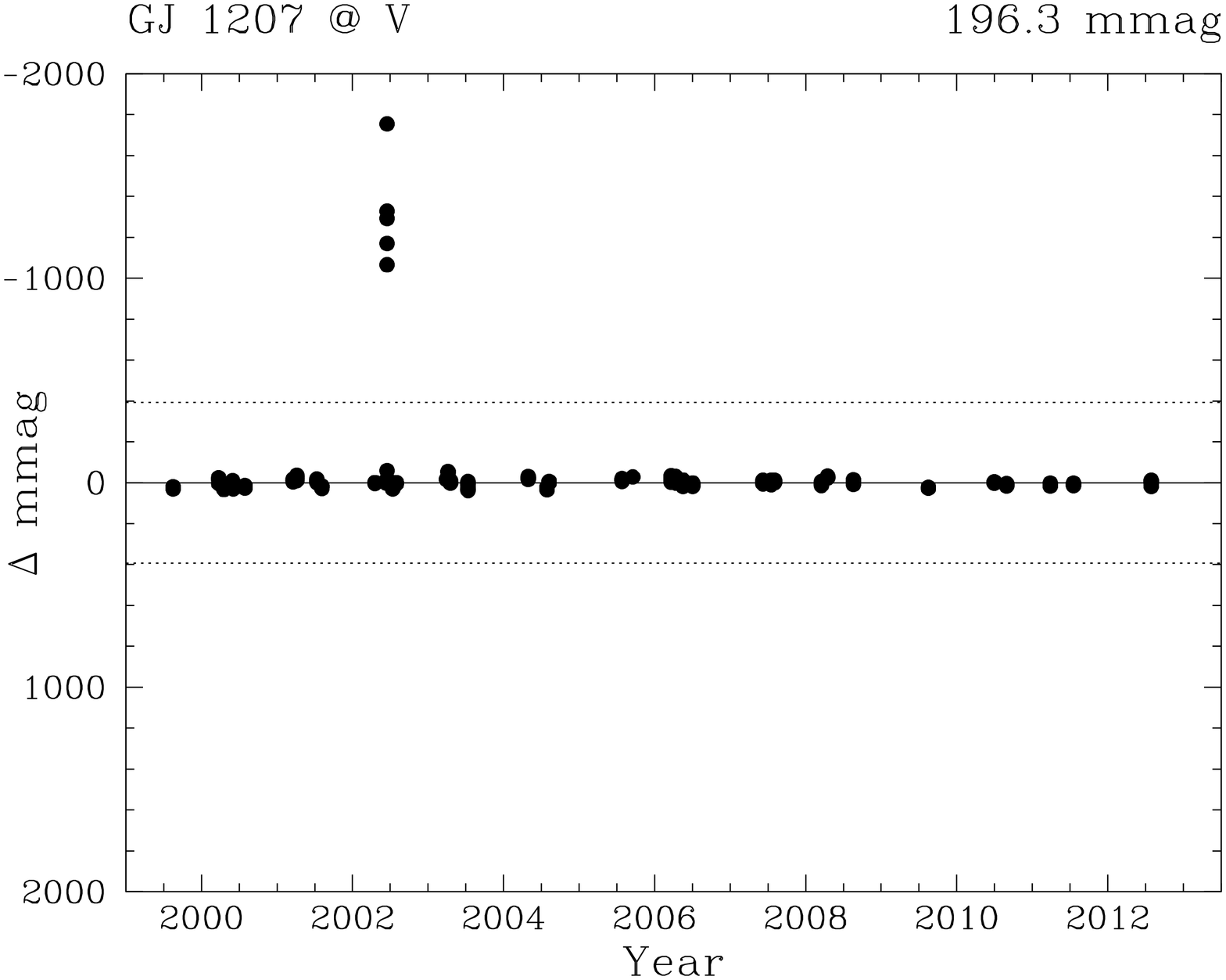}}
\hskip-10pt
{\includegraphics[scale=0.28,angle=270]{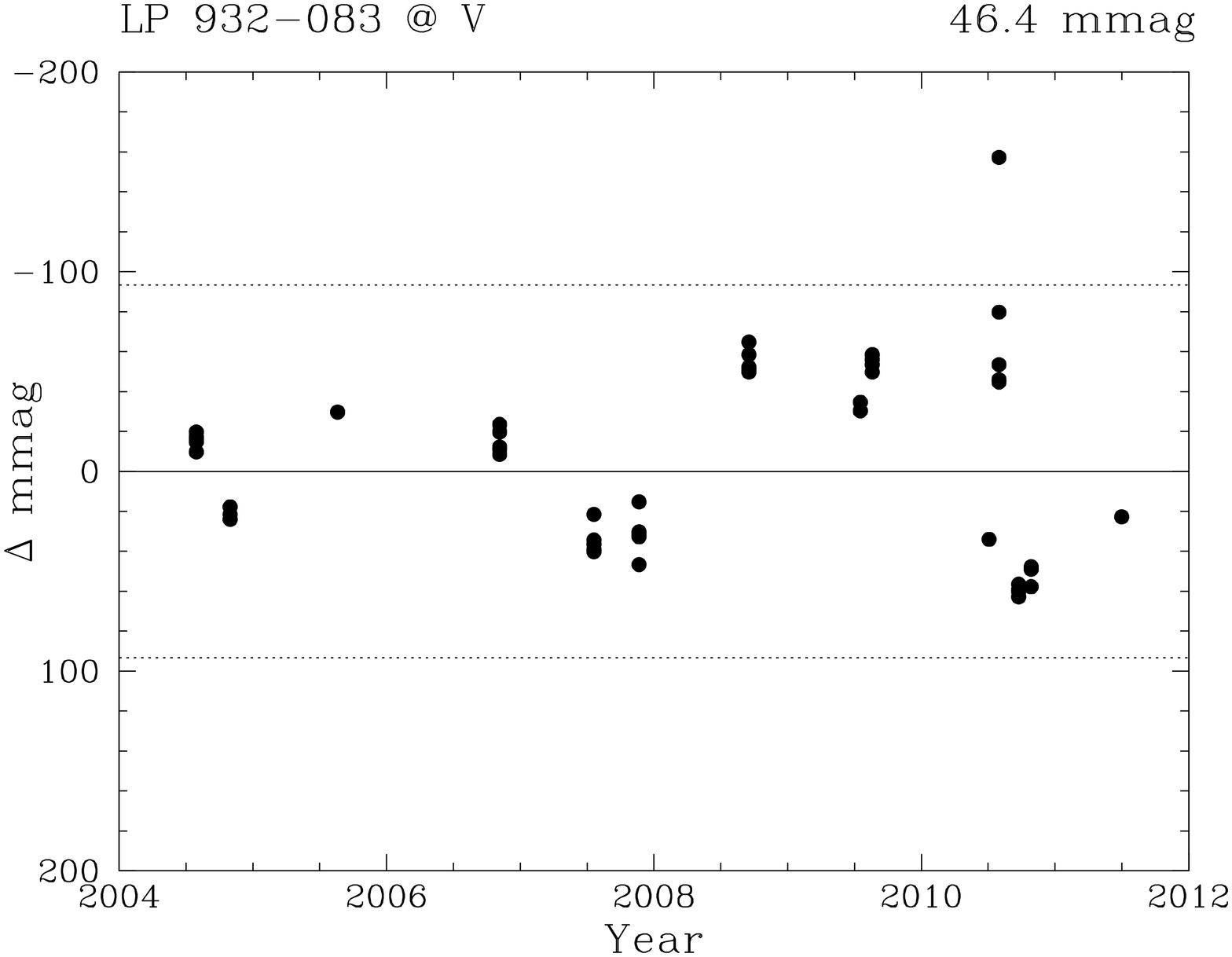}}
\hskip15pt
{\includegraphics[scale=0.28,angle=270]{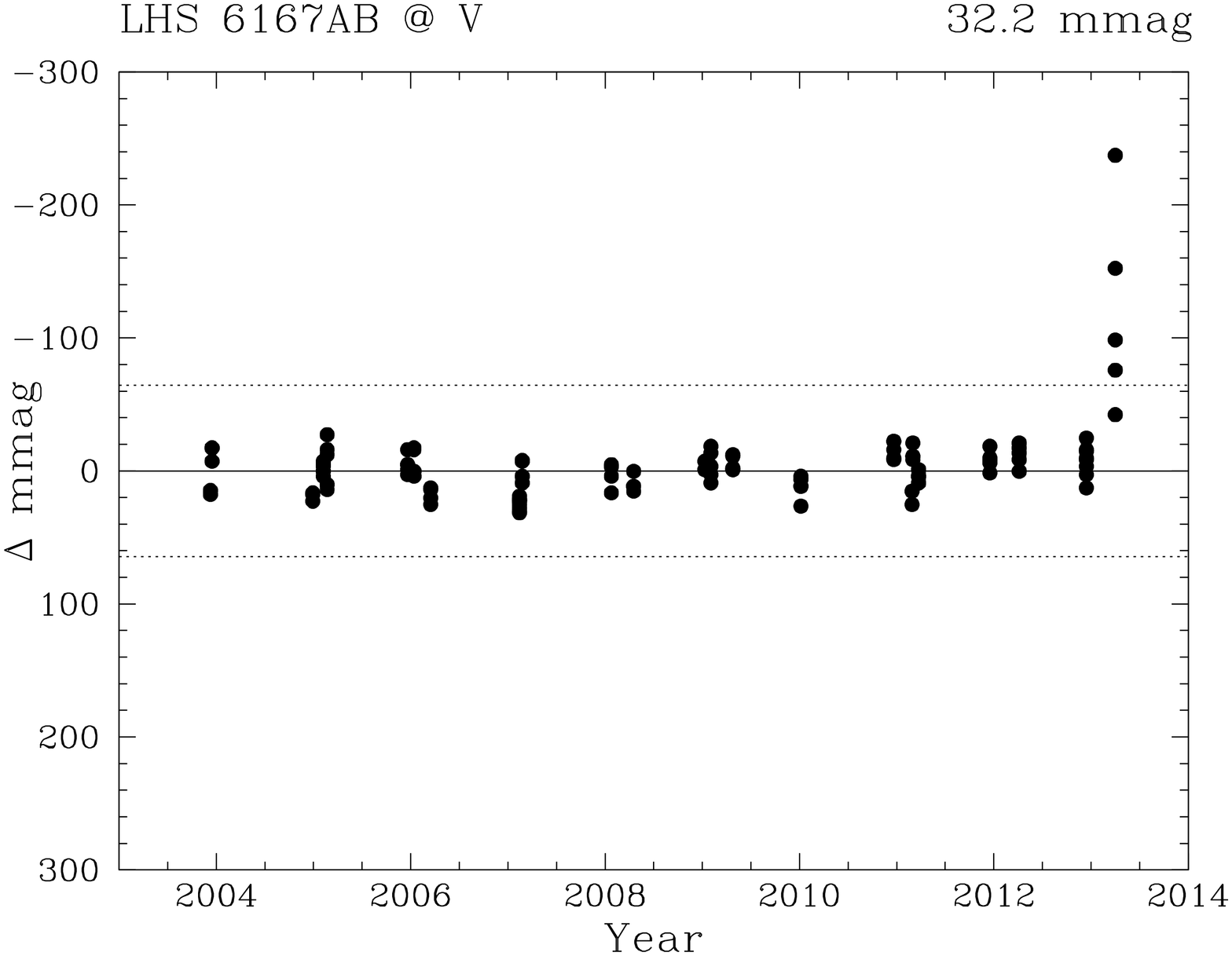}}
\hskip-10pt
{\includegraphics[scale=0.28,angle=270]{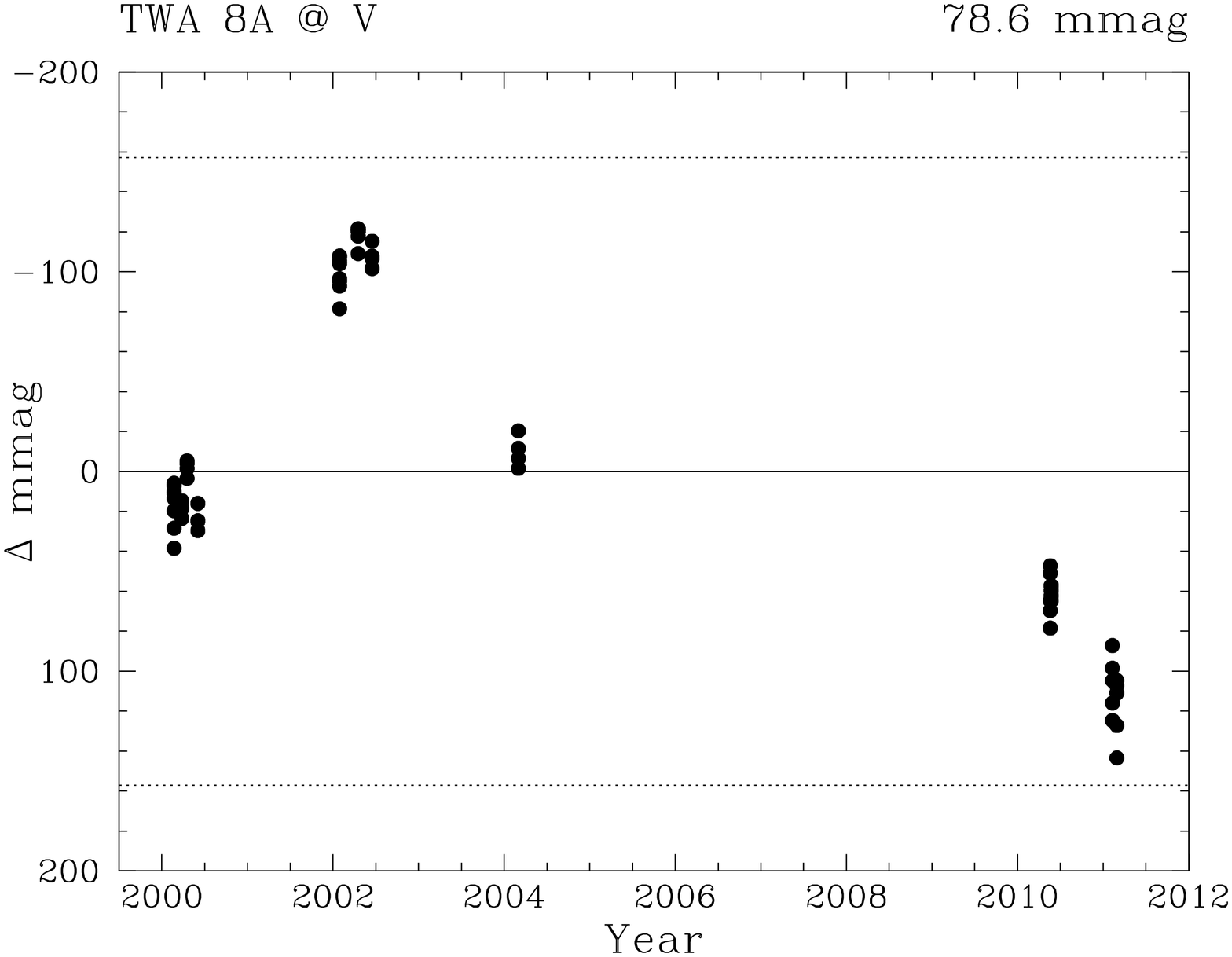}}
\hskip15pt
{\includegraphics[scale=0.28,angle=270]{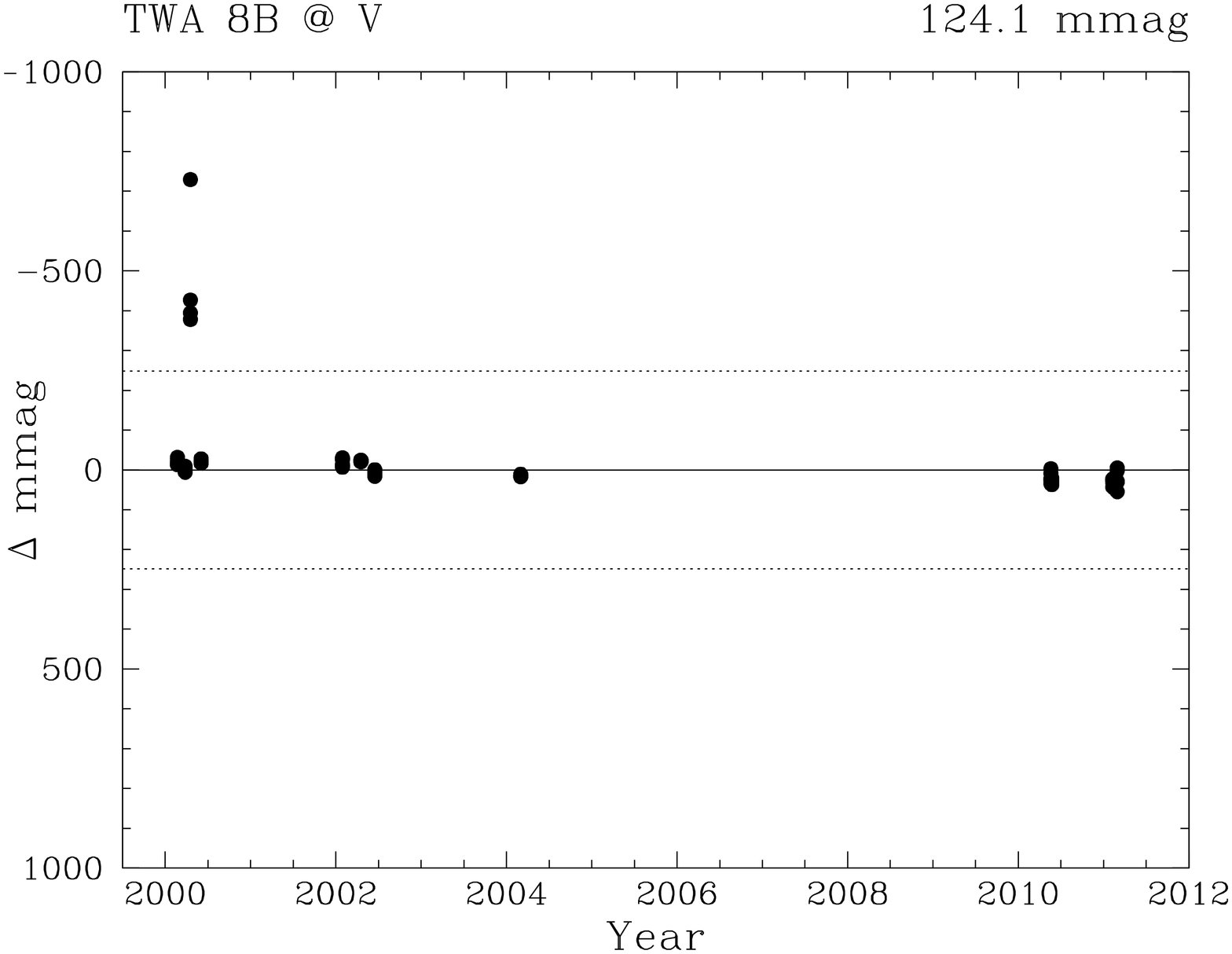}}

\vskip20pt

\caption{See Figure 6 caption for plot layout details. The three left
  panels show data for stars that appear to show evidence of spots,
  whereas the three right panels show stars with obvious flare
  events.}

\end{center}
\end{figure}



\begin{thebibliography}{}

\bibitem[Baraffe \& Chabrier(1996)]{1996ApJ...461L..51B} Baraffe, I.,
\& Chabrier, G.\ 1996, \apjl, 461, L51

\bibitem[Bessel(1990)]{1990A&AS...83..357B} Bessel, M.~S.\ 1990,
\aaps, 83, 357

\bibitem[Boyd et al.(2011b)]{2011AJ....142...92B} Boyd, M.~R., Henry,
T.~J., Jao, W.-C., Subasavage, J.~P., \& Hambly, N.~C.\ 2011b, \aj,
142, 92

\bibitem[Boyd et al.(2011a)]{2011AJ....142...10B} Boyd, M.~R., Winters,
J.~G., Henry, T.~J., et al.\ 2011a, \aj, 142, 10

\bibitem[Chabrier \& Baraffe(1997)]{1997A&A...327.1039C} Chabrier, G.,
\& Baraffe, I.\ 1997, \aap, 327, 1039

\bibitem[Ciardi et al.(2011)]{2011AJ....141..108C} Ciardi, D.~R., von
Braun, K., Bryden, G., et al.\ 2011, \aj, 141, 108

\bibitem[Davison et al.(2014)]{ 2014AJ....147...26D} Davison, C.~L. et
al.\ 2014, \aj, 147, 26

\bibitem[Dieterich et al.(2014)]{2014AJ....147...94D} Dieterich,
S.~B., Henry, T.~J., Jao, W.-C., et al.\ 2014, \aj, 147, 94

\bibitem[Finch et al.(2007)]{2007AJ....133.2898F} Finch, C.~T., Henry,
T.~J., Subasavage, J.~P., Jao, W.-C., \& Hambly, N.~C.\ 2007, \aj,
133, 2898

\bibitem[Franz et al.(1997)]{1997AAS...191.9302F} Franz, O.~G.,
Wasserman, L.~H., Henry, T.~J., et al.\ 1997, Bulletin of the American
Astronomical Society, 29, 1361

\bibitem[Gomes da Silva et al.(2012)]{2012A&A...541A...9G} Gomes da
Silva, J., Santos, N.~C., Bonfils, X., et al.\ 2012, \aap, 541, A9

\bibitem[Goulding et al.(2012)]{2012MNRAS.427.3358G} Goulding, N.~T.,
Barnes, J.~R., Pinfield, D.~J., et al.\ 2012, \mnras, 427, 3358

\bibitem[Graham(1982)]{1982PASP...94..244G} Graham, J.~A.\ 1982,
\pasp, 94, 244

\bibitem[Hambly et al.(2004)]{2004AJ....128..437H} Hambly, N.~C.,
Henry, T.~J., Subasavage, J.~P., Brown, M.~A., \& Jao, W.~C.\ 2004,
\aj, 128, 437 

\bibitem[Hartman et al.(2011)]{2011AJ....141..166H} Hartman, J.~D.,
Bakos, G.~{\'A}., Noyes, R.~W., et al.\ 2011, \aj, 141, 166

\bibitem[Hawley et al.(2014)]{2014ApJ...797..121H} Hawley, S.~L. et
al.\ 2014, \apj, 797, 121

\bibitem[Henry et al.(1999)]{1999ApJ...512..864H} Henry, T.~J., Franz,
O.~G., Wasserman, L.~H., et al.\ 1999, \apj, 512, 864

\bibitem[Henry et al.(2006)]{2006AJ....132.2360H} Henry, T.~J., Jao,
W.~C., Subasavage, J.~P., Beaulieu, T.~D., Ianna, P.~A., Costa, E., \&
Mendez, R.~A. \ 2006, \aj, 132, 2360 

\bibitem[Henry et al.(1994)]{1994AJ....108.1437H} Henry, T.~J.,
Kirkpatrick, J.~D., \& Simons, D.~A.\ 1994, \aj, 108, 1437 

\bibitem[Henry et al.(2004)]{2004AJ....128.2460H} Henry, T.~J.,
Subasavage, J.~P., Brown, M.~A., Beaulieu, T.~D., Jao, W.~C., \&
Hambly, N.~C.\ 2004, \aj, 128, 2460 

\bibitem[Honeycutt(1992)]{1992PASP..104..435H} Honeycutt, R.~K.\ 1992,
\pasp, 104, 435

\bibitem[Irwin et al.(2011)]{2011ApJ...727...56I} Irwin, J., Berta,
Z.~K., Burke, C.~J., et al.\ 2011, \apj, 727, 56

\bibitem[Jao et al.(2005)]{2005AJ....129.1954J} Jao, W.-C., Henry,
T.~J., Subasavage, J.~P., Brown, M.~A., Ianna, P.~A., Bartlett, J.~L.,
Costa, E., \& M{\'e}ndez, R.~A.\ 2005, \aj, 129, 1954 

\bibitem[Jao et al.(2011)]{2011AJ....141..117J} Jao, W.-C., Henry,
T.~J., Subasavage, J.~P., et al.\ 2011, \aj, 141, 117 

\bibitem[Jao et al.(2014)]{2014AJ....147..21J} Jao, W.-C., Henry,
T.~J., Subasavage, J.~P., et al.\ 2014, \aj, 147, 21 

\bibitem[Koen et al.(2010)]{2010MNRAS.403.1949K} Koen, C., Kilkenny, D., 
van Wyk, F., \& Marang, F.\ 2010, \mnras, 403, 1949 

\bibitem[Landolt(1992)]{1992AJ....104..372L} Landolt, A.~U.\ 1992,
\aj, 104, 372

\bibitem[Landolt(2007)]{2007AJ....133.2502L} Landolt, A.~U.\ 2007,
\aj, 133, 2502

\bibitem[Lurie et al.(2014)]{2014AJ....148...91L} Lurie, J.~C., Henry,
T.~J., Jao, W.-C., et al.\ 2014, \aj, 148, 91

\bibitem[Mamajek et al.(2013)]{2013AJ....146..154M} Mamajek, E.~E.,
Bartlett, J.~L., Seifahrt, A., et al.\ 2013, \aj, 146, 154

\bibitem[Montagnier et al.(2006)]{2006A&A...460L..19M} Montagnier, G.,
S{\'e}gransan, D., Beuzit, J.-L., et al.\ 2006, \aap, 460, L19

\bibitem[Riedel et al.(2014)]{2014AJ....147...85R} Riedel, A.~R.,
Finch, C.~T., Henry, T.~J., et al.\ 2014, \aj, 147, 85

\bibitem[Riedel et al.(2011)]{2011AJ....142..104R} Riedel, A.~R.,
Murphy, S.~J., Henry, T.~J., et al.\ 2011, \aj, 142, 104 

\bibitem[Riedel et al.(2010)]{2010AJ....140..897R} Riedel, A.~R.,
Subasavage, J.~P., Finch, C.~T., et al.\ 2010, \aj, 140, 897 

\bibitem[Robertson et al.(2013)]{2013ApJ...764....3R} Robertson, P.,
Endl, M., Cochran, W.~D., \& Dodson-Robinson, S.~E.\ 2013, \apj, 764,
3

\bibitem[Subasavage et al.(2005a)]{2005AJ....129..413S} Subasavage,
J.~P., Henry, T.~J., Hambly, N.~C., Brown, M.~A., \& Jao, W.-C.\
2005a, \aj, 129, 413

\bibitem[Subasavage et al.(2005b)]{2005AJ....130.1658S} Subasavage,
J.~P., Henry, T.~J., Hambly, N.~C., et al.\ 2005b, \aj, 130, 1658

\bibitem[Subasavage et al.(2009)]{2009AJ....137.4547S} Subasavage,
J.~P., Jao, W.-C., Henry, T.~J., et al.\ 2009, \aj, 137, 4547 

\bibitem[Walkowicz et al.(2011)]{2011AJ....141...50W} Walkowicz,
L.~M., Basri, G., Batalha, N., et al.\ 2011, \aj, 141, 50

\bibitem[Weis(1994)]{1994AJ....107.1135W} Weis, E.~W.\ 1994, \aj, 107,
1135

\bibitem[Weis(1996)]{1996AJ....112.2300W} Weis, E.~W.\ 1996, \aj, 112,
2300

\bibitem[Winters et al.(2011)]{2011AJ....141...21W} Winters, J.~G.,
Henry, T.~J., Jao, W.-C., Subasavage, J.~P., Finch, C.~T., \& Hambly,
N.~C.\ 2011, \aj, 141, 21

\bibitem[Winters et al.(2015)]{2015AJ....149....5W} Winters, J.~G.,
Henry, T.~J., Lurie, J.C., et al.\ 2015, \aj, 149, 5

\end{thebibliography}
\end{document}